\documentclass[letterpaper]{article} % DO NOT CHANGE THIS
\usepackage{aaai23}  % DO NOT CHANGE THIS
\usepackage{times}  % DO NOT CHANGE THIS
\usepackage{helvet}  % DO NOT CHANGE THIS
\usepackage{courier}  % DO NOT CHANGE THIS
\usepackage[hyphens]{url}  % DO NOT CHANGE THIS
\usepackage{graphicx} % DO NOT CHANGE THIS
\def\UrlFont{\rm}  % DO NOT CHANGE THIS
\usepackage{natbib}  % DO NOT CHANGE THIS AND DO NOT ADD ANY OPTIONS TO IT
\usepackage{caption} % DO NOT CHANGE THIS AND DO NOT ADD ANY OPTIONS TO IT
\usepackage{algorithm}
\usepackage{algorithmic}

\usepackage{newfloat}
\usepackage{listings}
\DeclareCaptionStyle{ruled}{labelfont=normalfont,labelsep=colon,strut=off} % DO NOT CHANGE THIS
\floatstyle{ruled}
\newfloat{listing}{tb}{lst}{}
\floatname{listing}{Listing}
\usepackage{amsmath}

\usepackage{amssymb}
\usepackage[capitalise]{cleveref}
\usepackage{multirow}
\usepackage{graphicx}
\usepackage{booktabs}% toprule, midrule

\DeclareMathOperator*{\E}{\mathbb{E}}

\newcommand{\rpm}{\sbox0{$1$}\sbox2{$\scriptstyle\pm$}
  \raise\dimexpr(\ht0-\ht2)/2\relax\box2 }

\title{Towards Reliable Item Sampling for Recommendation Evaluation}
\author{
       Dong Li\textsuperscript{\rm 1},
    Ruoming Jin\textsuperscript{\rm 1},
    Zhenming Liu\textsuperscript{\rm 2},
    Bin Ren\textsuperscript{\rm 2},
    Jing Gao\textsuperscript{\rm 3},
    Zhi Liu\textsuperscript{\rm 3}
   }
   \affiliations {
       \textsuperscript{\rm 1} Kent State University,\\
       \textsuperscript{\rm 2}  College of William \& Mary,\\
       \textsuperscript{\rm 3}   iLambda,\\
       \{dli12, rjin1\}@kent.edu,
       \{zliu, bren\}@cs.wm.edu,
       \{jgao, zliu\}@ilambda.com
}

\begin{document}

\maketitle

\begin{abstract}
Since Rendle and Krichene argued that commonly used sampling-based evaluation metrics are ``inconsistent'' with respect to the global metrics (even in expectation), there have been a few studies on the sampling-based recommender system evaluation. Existing methods try either mapping the sampling-based metrics to their global counterparts or more generally, learning the empirical rank distribution to estimate the top-$K$ metrics. 
However, despite existing efforts, there is still a lack of rigorous theoretical understanding of the proposed metric estimators, and the basic item sampling also suffers from the ``blind spot'' issue, i.e., estimation accuracy to recover the top-$K$ metrics when $K$ is small can still be rather substantial. 
In this paper, we provide an in-depth investigation into these problems and make two innovative contributions. First, we propose a new item-sampling estimator that explicitly optimizes the error with respect to the ground truth, and theoretically highlights its subtle difference against prior work. Second, we propose a new adaptive sampling method that aims to deal with the ``blind spot'' problem and also demonstrate the
expectation-maximization (EM) algorithm can be generalized for such a setting. 
Our experimental results confirm our statistical analysis and the superiority of the proposed works. 
This study helps lay the theoretical foundation for adopting item sampling metrics for recommendation evaluation and provides strong evidence for making item sampling a powerful and reliable tool for recommendation evaluation. 
\end{abstract}
\section{Introduction}
As personalization and recommendation continue to play an integral role in the emerging AI-driven economy \cite{fayyaz2020recommendation,zhao2021recbole,graph@sigir22,jin2021towards,GREASE}, proper and rigorous evaluation of recommendation models become increasingly important in recent years for both academic researchers and industry practitioners~\cite{Gruson2019,cremonesi2011comparative,dacrema2019troubling,rendle2019evaluation}. 
Particularly, ever since \citet{Krichene20@KDD20,DBLP:journals/cacm/KricheneR22}  pointed out the ``inconsistent'' issue of  item-sampling based evaluation of commonly used (top-$K$) evaluation metrics, such as Recall (Hit-Ratio)/Precision, Average Precision (AP) and Normalized Discounted Cumulative Gain (NDCG), (other than AUC), it has emerged as a major controversy being hotly debated among recommendation community. 

Specifically, the item-sampling strategy calculates the top-$K$ evaluation metrics using only a small set of item samples~\cite{Koren08, CremonesiKT@10,he2017neural, ebesu2018collaborative,HuSWY18,krichene2018efficient,wang2018explainable,YangBGHE18,YangCXWB18}. \citet{Krichene20@KDD20,DBLP:journals/cacm/KricheneR22} show that the top-$K$ metrics based on the samples differ from the global metrics using all the items.  They suggested a cautionary use (avoiding if possible) of the sampled metrics for recommendation evaluation. 
Due to the ubiquity of sampling methodology, it is not only of theoretical importance but also of practical interest to understand item-sampling evaluation. Indeed, since the number of items in any real-world recommendation system is typically quite large (easily in the order of tens of millions), efficient model evaluation based on item sampling can be very useful for recommendation researchers and practitioners.  

To address the discrepancy between item sampling results and the exact top-$K$ recommendation evaluation, \citet{Krichene20@KDD20} have proposed a few estimators in recovering global top-$K$ metrics from sampling. Concurrently, \citet{Li@KDD20} showed that for the top-$K$ Hit-Ratios (Recalls) metric, there is an (approximately) linear relationship between the item-sampling top-$k$ and the global top-$K$ metrics ($K=f(k)$, where $f$ is approximately linear). In another recent study, \citet{Jin@AAAI21} developed solutions based on MLE (Maximal Likelihood Estimation) and ME (Maximal Entropy)  to learn the {\em empirical rank distribution}, which is then used to estimate global top-$K$ metrics.

Despite these latest works on item-sampling estimation~\cite{Li@KDD20, Krichene20@KDD20, Jin@AAAI21},  there remain some major gaps in making item-sampling reliable and accurate for top-$K$ metrics. Specifically, the following important problems are remaining open:  

\emph{(i)} What is the optimal estimator given the basic item sampling?  All the earlier estimation methods do not establish any optimality results with respect to the estimation errors~\cite{Krichene20@KDD20,Jin@AAAI21}. 

\emph{(ii)} What can we do for the problem of the basic item sampling, which appears to have a  fundamental limitation that prevents us from recovering  the global rank distributions accurately?
For the offline recommendation evaluation, we typically are interested in the top-ranked items and top-$K$ metrics, when $K$ is relatively small, say less than $50$. 
However, the current item sampling seems to have a ``blind spot'' for the top-rank distribution. For example, when there are $n = 100$ samples and $N = 10k$, the estimation granularity is only at around 1\% ($1/n$) level~\cite{Krichene20@KDD20,Li@KDD20}. We can only infer that the top items in the samples are top 1\% (top 100) in the global rank, while we can not further tell whether the top items in the sample set are in, for example, top-50, without increasing the sampling size. Given this, even with the best estimator for the item sampling, we may still not be able to provide accurate results for the top-$K$ metrics.   
A remedy is increasing the sampling size, but it can significantly increase the estimation cost too, limiting the benefits of item sampling. 
Can we sample the items in a more intelligent manner to circumvent the ``blind spot'' while keeping estimation cost low (and the sample size small)? To address the above open questions, we make the following contributions in this paper: 

\begin{itemize}
        \item We derive an optimal item-sampling estimator and highlight subtle differences from the BV estimators derived by \citet{Krichene20@KDD20}, and point out the potential issues of BV estimator because it fails to link the user population size with the estimation variance.  To the best of our knowledge, this is the first estimator that directly optimizes the estimation errors. 

        \item  We address the limitation of the current item sampling approaches by proposing a new  adaptive sampling method. This provides a simple and effective remedy that helps avoid the sampling ``blind spot'', and significantly improves the accuracy of the estimated metrics with low sample complexity. 

            \item We perform a thorough experimental evaluation of the proposed  item-sampling estimator and the new adaptive sampling method. The experimental results further confirm the statistical analysis and the superiority of newly proposed estimators.
    \end{itemize}

Our results help lay the theoretical foundation for adopting item sampling metrics for recommendation evaluation and offer a simple yet effective new adaptive sampling approach to help recommendation practitioners and researchers to apply item sampling-based approaches to speed up offline evaluation. The following is organized: \Cref{sec:problem} introduces the item sampling based top-$K$ evaluation framework and reviews the related work (in \Cref{sec:app_related}); \Cref{sec:newestimator} introduces the new optimal estimator minimizing its mean squared error with respect to the ground-truth; 
\Cref{sec:adaptive} presents the new adaptive sampling and estimation method; 
\Cref{sec:exp} discusses the experimental results; 
and finally, \Cref{sec:conc} concludes the paper.

\section{Overview}\label{sec:problem}

\cref{tab:notations} (in \cref{sec:note}) highlights the key notations for evaluating recommendation algorithms used throughout the paper. Given $M$ users and $N$ items.
To evaluate the quality of recommender models, each testing user $u$ hides an already clicked (or so-called target) item $i_u$, and compares it with the rest of the items, derives a rank $R_u$. The recommendation  model is considered to be effective if it ranks $i_u$ at the top (small $R_u$). 
Formally, given a recommendation model, a metric function (denoted as a metric $\mathcal{M}$) maps each rank $R_u$ to a real-valued score, and then averages over all the users in the test set: 
\begin{small}
\begin{equation}\label{eq:metric_0}
T =\frac{1}{M}\sum_{u=1}^M {\mathcal M}(R_u) 
\end{equation}
\end{small}
And the corresponding top-K evaluation metric:
\begin{small}
\begin{equation}\label{eq:topk_metric}
T=\frac{1}{M}\sum_{u=1}^M {\bf 1}_{R_u\leq K} \cdot {\mathcal M}(R_u) 
\end{equation}
\end{small}
where $\mathbf{1}_x=1$ if $x$ is True, 0 otherwise.

The commonly used function ${\mathcal M}$ of evaluation metrics~\cite{Krichene20@KDD20} are Recall, Precision, AUC, NDCG, and AP. For example: 
{\small 
\begin{equation}\label{eq:recall_topk_metric}
\begin{split}
&Recall@K=\frac{1}{M} \sum_{u=1}^M {\bf 1}_{R_u\leq K}
\end{split}
\end{equation}
}

\subsection{Item-Sampling Top-K Evaluation}
In the item-sampling-based top-K evaluation scenario, 
for a given user $u$ and his/her relevant item $i_u$, another $n - 1$ items from the entire item set $I$ are sampled. The union of sampled items and $i_u$ is $I_u$ ($i_u \in I_u$, $|I_u|=n$). The recommendation model then returns the rank of $i_u$ among $I_u$, denoted as $r_u$  (again, $R_u$ is the rank against the entire set of items $I$).

Given this, a list of studies  ~\cite{Koren08, he2017neural} simply replaces $R_u$ with $r_u$ for (top-$K$) evaluation. Sampled evaluation metric/performance denoted as : 

\begin{equation}\label{eq: empirical_m}
T_S\triangleq\frac{1}{M}\sum_{u=1}^M {\bf 1}_{r_u\leq K}\cdot {\mathcal M}(r_u) 
\end{equation}

It's obvious that $r_u$ and $R_u$ differ substantially, for example, $r_u \in [1,n]$ whereas $R_u \in [1,N]$. Therefore, for the same $K$, the item-sampling top-K metrics and the global top-K metrics correspond to distinct measures (no direct relationship): $T  \neq T_S$ ($Recall@K \neq Recall_S@K$). 
This problem is highlighted  in~\cite{Krichene20@KDD20, rendle2019evaluation}, referring to these two metrics being {\em inconsistent}.
From the perspective of statistical inference, the basic sampling-based top-$K$ metric $T_S@K$  is not a {reasonable} or good {\em estimator} ~\cite{theorypoint} of $T@K$. 

\citet{Li@KDD20} showed that for some of the most commonly used metrics, the top-K Recall/HitRatio, there is a  mapping function $f$ (approximately linear), such that 
$Recall@f(k) \approx {Recall}_S@k$. Thus, they give an intuitive explanation on how to look at a sampling top-$K$ metrics (on Recall) linking it the same global metric but at a different rank/location.

There are two recent works (see \cref{sec:app_related}) studying the general metric estimation problem based on the item sampling metrics. Specifically, given the sampling ranked results in the test set, $\{r_u\}^{M}_{u=1}$, how to infer/approximate the $T$ from \cref{eq:metric_0} or more commonly \cref{eq:topk_metric}, without the knowledge $\{R_u\}_{u=1}^M$?

\noindent{\bf Two Problems:}
We consider two fundamental (open) questions for item-sampling estimation: 
1) What is the optimal estimator following for \cite{Krichene20@KDD20}? (Their methods do not directly target minimizing  the estimation errors). 
2) By solving the first problem and the MLE method from \cite{Jin@AAAI21}, we observe the best effort using the basic item sampling still fails to recover accurately on the global top-$K$ metrics when $K$ is small, as well as the rank distribution for the top spots. Those ``blind spots'' seem to stem from the inherent ({\em information}) limitation of item sampling methods, not from the estimators. How can we effectively address such item-sampling limitations?
Next, we will introduce methods to address these two problems.

\section{New Estimator for Item-Sampling}\label{sec:newestimator}
In this section, we introduce a new estimator which aims to directly minimize the expected errors between the item-sampling-based top-$K$ metrics and the global top-$K$ metrics. Here, we consider a similar strategy as ~\cite{Krichene20@KDD20} though our objective function is different and aims to explicitly minimize the expected error. 
We aim to search for a {\em sampled metric} $\widehat{\mathcal M}(r)$  to approach $\widehat{T}\approx T$: 
\begin{small}
\begin{equation*}
\begin{split}
    \widehat{T}&=\sum_{r=1}^n \tilde{P}(r) \cdot \widehat{\mathcal M}(r) = \frac{1}{M}\sum_{u=1}^M \widehat{\mathcal M}(r_u) \\ &\approx  \frac{1}{M}\sum_{u=1}^M {\mathcal M}(R_u) = \sum_{R=1}^N \tilde{P}(R)\cdot {\mathcal M}(R) = T 
    \end{split}
\end{equation*}
\end{small}
where $\tilde{P}(r) = \frac{1}{M}\sum\limits_{r=1}^M\mathbf{1}_{r_u=r}$ is the empirical sampled rank distribution
and $\widehat{\mathcal M}(r)$ is the adjusted discrete metric function. An immediate observation is: 
\begin{equation}\label{eq:t_et}
    \begin{split}
        \E \widehat{T}&=\sum_{r=1}^n \E [\tilde{P}(r)]\cdot \widehat{\mathcal M}(r) = \sum_{r=1}^n {P}(r) \cdot  \widehat{\mathcal M}(r)
    \end{split}
\end{equation}

Following the classical statistical inference~\cite{casella2002statistical}, the optimality of an estimator is measured by Mean Squared Error (more derivation in \cref{sec:mse_re}):

\begin{equation}\label{eq:last0}
\E[\widehat{T}-\E T]^2 =        \E[\widehat{T}-\sum_{R=1}^N P(R) \mathcal{M}(R)]^2 \\
\end{equation}
\begin{equation*}    
\begin{split}
  =&  \Big(\sum_{r=1}^n \sum_{R=1}^N P(r|R)P(R) \widehat{\mathcal M}(r)-\sum_{R=1}^N P(R) \mathcal{M}(R)\Big)^2  \nonumber \\
  +& \E[\sum_{r=1}^n \sum_{R=1}^N \tilde{P}(r|R)P(R) \widehat{\mathcal M}(r)- \sum_{r=1}^n \sum_{R=1}^N  P(r|R)P(R) \widehat{\mathcal M}(r)]^2 
    \end{split}
\end{equation*}

Remark that $\tilde{P}(r|R)$ is the empirical conditional sampling rank distribution given a global rank $R$. 
We next use Jensen's inequality to bound the first term in (\cref{eq:last0}). Specifically, we may treat $\sum_{r=1}^n P(r|R)  \widehat{\mathcal M}(r)- \mathcal{M}(R)$ as a random variable and use $(\E X)^2 \leq \E X^2$ to obtain
\begin{small}
\begin{equation*}
    \begin{split}
         &\Big(\sum_{r=1}^n \sum_{R=1}^N P(r|R)P(R) \widehat{\mathcal M}(r)-\sum_{R=1}^N P(R) \mathcal{M}(R)\Big)^2   \\
         &\leq 
    \sum_{R=1}^N P(R) \Big(\sum_{r=1}^n P(r|R)  \widehat{\mathcal M}(r)- \mathcal{M}(R)\Big)^2
    \end{split}
\end{equation*}
\end{small}

% \begin{scriptsize}
% \begin{align*}
%     \Big(\sum_{r=1}^n \sum_{R=1}^N P(r|R)P(R) \widehat{\mathcal M}(r)-\sum_{R=1}^N P(R) M(R)\Big)^2   \leq 
%     \sum_{R=1}^N P(R) \Big(\sum_{r=1}^n P(r|R)  \widehat{\mathcal M}(r)- M(R)\Big)^2
% \end{align*}
% \end{scriptsize}
Therefore, we have 

\begin{small}
\begin{equation*}
    \begin{split}
         &\E[\widehat{T}-\sum_{R=1}^N P(R) \mathcal{M}(R)]^2 \\  
    &\leq  \underbrace{\sum_{R=1}^N P(R) \Big\{ \Big(\sum_{r=1}^n P(r|R)  \widehat{\mathcal M}(r)- \mathcal{M}(R)\Big)^2}_{\mathcal L_1} \\
    &+ \underbrace{\E[\sum_{r=1}^n \tilde{P}(r|R) \widehat{\mathcal M}(r)- \sum_{r=1}^n  P(r|R) \widehat{\mathcal M}(r)]^2 \Big\}}_{\mathcal L_2}.
    \end{split}
\end{equation*}
\end{small}

Let $\mathcal L = \mathcal L_1 + \mathcal L_2$, which gives an upper bound on the expected MSE. Therefore, our goal is to find $\widehat{\mathcal{M}}(r)$ to minimize $\mathcal L$. We remark that a seemingly innocent application of Jensen's inequality results in an optimization objective that possesses a range of interesting properties:

\noindent{\textbf{1. Statistical structure.}} The objective has a variance-bias trade-off interpretation, i.e., 
\begin{small}
       \begin{equation}
       \label{eq:L1}
\mathcal{L}_1=\sum\limits_{R=1}^N{P(R)\Big({\E(\widehat{\mathcal M}(r)|R) -\mathcal{M}(R)}\Big)^2} 
\end{equation}
\begin{equation}
        \label{eq:L2}
            \mathcal{L}_2 
 = \sum\limits_{R=1}^{N}{\frac{1}{M} Var(\widehat{\mathcal M}(r)|R)}
\end{equation}
\end{small}

where $\mathcal L_1$ can be interpreted as a bias term and $\mathcal L_2$ can be interpreted as a variance term. Note that while \citet{Krichene20@KDD20} also introduce a variance-bias tradeoff objective, their objective is constructed from heuristics and contains a hyper-parameter (that determines the relative weight between bias and variance) that needs to be tuned in an ad-hoc manner. Here, because our objective is constructed from direct optimization of the MSE, it is more principled and also removes dependencies on hyperparameters. See \cref{sec:analysisl2} for proving ~\cref{eq:L2} (~\cref{eq:L1} is trivial) and ~\cref{sec:closedform} for more comparison against estimators proposed in~\cite{Krichene20@KDD20}.

\noindent{\textbf{2. Algorithmic structure.}} while the objective is not convex, we show that the objective can be expressed in a compact manner using matrices and we can find the optimal solution in a fairly straightforward manner. In other words, Jensen's inequality substantially simplifies the computation at the cost of having a looser upper bound. See ~\cref{sec:closedform}. 

\noindent{\textbf{3. Practical performance.}} Our experiments also confirm that the new estimator is effective (~\cref{sec:exp}), which suggests that Jensen's inequality makes only inconsequential and moderate performance impact on the estimator's quality.  

\subsection{Analysis of ${\mathcal L}_2$}\label{sec:analysisl2}
To analyze ${\mathcal L}_2$, let us take a close look at $\tilde{P}(r|R)$. Formally, let $X_r$ be the random variable representing the number of items at rank $r$ in the item-sampling data whose original rank in the entire item set is $R$. Then, we rewrite  $\tilde{P}(r|R) = \frac{X_r}{M\cdot P(R)}$. Furthermore, it is easy to observe $(X_1, \cdots X_n)$ follows the multinomial distribution $Multi(P(1|R), \cdots,P(n|R))$ (See $P(r|R)$ defined in \cref{tab:notations}). We also have:   
\begin{equation}
    \begin{split}        
        &\E[X_r]=M\cdot P(R)\cdot P(r|R) \\ 
        &Var[X_r] = M\cdot P(R)\cdot  P(r|R)(1-P(r|R))
    \end{split}
\end{equation}

Next, let us define a new random variable $\mathcal{B} \triangleq \sum\limits_r^{n}{\widehat{\mathcal M}(r) X_r}$, which is the weighted sum of random variables under a multinomial distribution. According to  \cref{sec:app_linear}, its variance is give by:
\begin{equation*}
    \begin{split}
        &Var[\mathcal{B}] =  \E[\sum_{r=1}^n {X_r}\widehat{\mathcal M}(r)- \sum_{r=1}^n  {\E[X_r]} \widehat{\mathcal M}(r)]^2\\
        &=M\cdot P(R) \Big(\sum\limits_{r}^n{\widehat{\mathcal M}^2(r) P(r|R)}-\big(\sum\limits_{r}^n{\widehat{\mathcal M}(r) P(r|R)}\big)^2 \Big) 
    \end{split}
\end{equation*}
${\mathcal L}_2$ can be re-written (see \cref{sec:l2_re}) as:
\begin{equation}
    \mathcal{L}_2 = \sum\limits_{R=1}^{N}{\frac{1}{M} Var(\widehat{\mathcal M}(r)|R)} \nonumber
\end{equation}

\subsection{Closed Form Solution and its Relationship to Bias-Variance Estimator}\label{sec:closedform}
We can rewrite $\mathcal L$ as a matrix format (see \cref{sec:app_ls}) and optimize it corresponding to a constraint least square optimization and its solution:
\begin{equation}\label{eq:new_loss}
    \begin{split}
        \mathcal{L} = ||\sqrt{D}A\mathbf{x}-\sqrt{D}\mathbf{b}||^2_F+\frac{1}{M}||\sqrt{\Lambda}_1\mathbf{x}||^2_F -\frac{1}{M} ||A\mathbf{x}||^2_F
    \end{split}
\end{equation}
\begin{equation}\label{eq:mn_closed}
    \begin{split}
        \mathbf{x}=\Big(A^TDA-\frac{1}{M}A^TA+\frac{1}{M}\Lambda_1\Big)^{-1}A^TD\mathbf{b}
    \end{split}
\end{equation}
where $M$ is the number of users and $\mathcal{M}$ is metric function, $diagM(\cdot)$ is a diagonal matrix:
\begin{small}
\begin{equation*}\label{eq:new_parameters}
\begin{split}
\mathbf{x} &= \begin{bmatrix}
        \widehat{\mathcal M}(r=1)\\
        \vdots\\
        \widehat{\mathcal M}(r=n)
        \end{bmatrix}\quad \mathbf{b}=\begin{bmatrix}
        \mathcal{M}(R=1)\\
        \vdots\\
        \mathcal{M}(R=N)
        \end{bmatrix}\in\mathbb{R}^N\\
        A_{R,r} &= P(r|R)\in \mathbb{R}^{N\times n}\quad
        D = diagM\big(P(R)\big)\in \mathbb{R}^{N\times N}\\
        \Lambda_1 &= diagM\big(\sum\limits_{R=1}^{N}P(r|R)\big)\in \mathbb{R}^{n\times n}\\
\end{split}
\end{equation*}
\end{small}

\noindent{\bf Relationship to the BV Estimator:}
The bias-variance trade-off is given by \cite{Krichene20@KDD20}:
\begin{small}
\begin{equation*}\label{eq:eq_bv}
\begin{split}
       \mathcal{L}_{BV}
       & = \underbrace{\sum\limits_{R=1}^{N} P(R) (\E[\widehat{\mathcal M}(r)|R]-\mathcal{M}(R))^2}_{{\mathcal L}_1} \\&+ \underbrace{\sum\limits_{R=1}^{N} P(R) \cdot \gamma \cdot Var[\widehat{\mathcal M}(r)|R]}_{{\mathcal L}_2}  \nonumber
       \end{split}
\end{equation*}
\end{small}

We observe the main difference between the BV and our new estimator is on the ${\mathcal L}_2$ components (variance components): for our estimator, each $Var[\widehat{\mathcal M}(r)|R]$ is regularized by $1/M$ ($M$ is the number of testing users), where in BV, this term is regularized by $P(R) \gamma$. Our estimator reveals that as the number of users increases, the variance in the ${\mathcal L}_2$ components will continue to decrease, whereas the BV estimator does not consider this factor. Thus, as the user size increases, BV estimator still needs to deal with ${\mathcal L}_2$ or has to manually adjust $\gamma$. 

Finally, both BV and the new estimator rely on prior distribution $P(R)$, which is unknown. In \cite{Krichene20@KDD20}, the uniform distribution is used for the estimation purpose. In this paper, we propose to leverage the latest approaches in ~\cite{Jin@AAAI21} which provide a more accurate estimation of $P(R)$ for this purpose. The experimental results in \cref{sec:exp} will confirm the validity of using such distribution  estimations.

\section{Adaptive Item-Sampling Estimator}
\label{sec:adaptive}

\begin{figure}
    \centering
    \includegraphics[width = 0.8\linewidth]{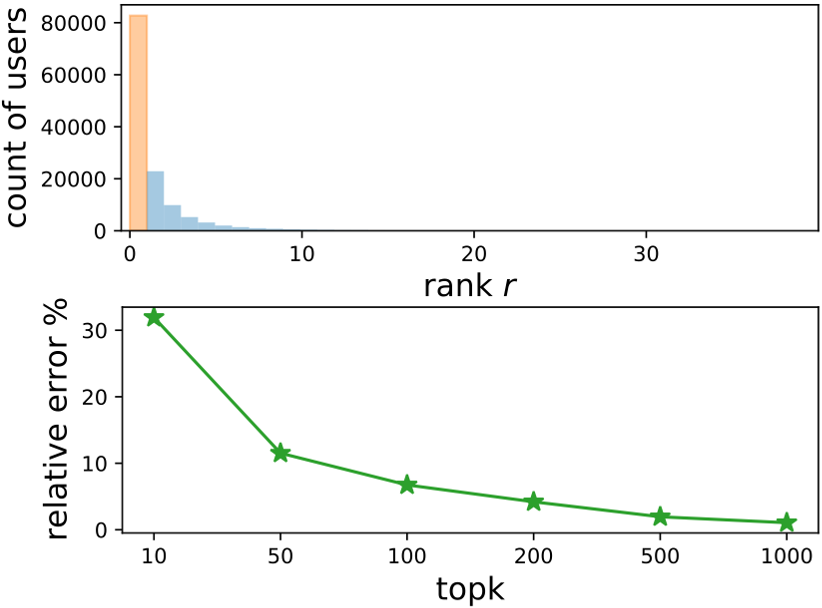}
    \caption{Top is distribution of $r_u$ with sample set size $n=100$. Rank $r=1$ is highlighted. The bottom is the relative error of MLE estimator for different top-$K$. The result is obtained by EASE model \cite{Steck_2019} over ml-20m dataset. }
    \label{fig:dist_ru}
\end{figure}

\subsection{Blind Spot and Adaptive Sampling}

In recommendation, top-ranked items are vital, thus it's more crucial to obtain an accurate estimation for these top items. However current sampling approaches treat all items equally and particularly have difficulty in recovering the global top-$K$ metrics when K is small. At top of \cref{fig:dist_ru}, we plot the distribution of target items' rank in the sample set and observe that most target items rank top 1 (highlighted in red). This could lead to the "blind spot" problem - when $K$ gets smaller, the estimation of basic estimators is more inaccurate (see bottom of \cref{fig:dist_ru}). Intuitively, when $r_u=1$, it does not mean its global rank $R_u$ is $1$, instead, its expected global rank may be around $100$ (assuming $N=10K$ and sample set size $n=100$).
And the estimation granularity is only at around 1\% ($1/n$) level. This blind spot effect brings a big drawback for current estimators.

Based the on the above discussion, we propose an adaptive sampling strategy, which increases acceptable test sample size for users whose target item ranks top (say $r_u=1$) in the sampled data. When $r_u = 1$, we continue doubling the sample size until $r_u\neq 1$ or until the sample size reaches a predetermined ceiling. See~\cref{alg:adp} (and detailed explanation in \cref{sec:adp_sampling}). The benefits of this adaptive strategy are two folds: \textit{high granularity}, with more items sampled, the counts of $r_u = 1$ shall reduce, which could further improve the estimating accuracy; \textit{efficiency}, we iteratively sample more items for users whose $r_u=1$ and the empirical experiments (\cref{tab:ndcg_adaptive}) confirm that small average adaptive sample size (compared to uniform sample size) is able to achieve significantly better performance.

\begin{algorithm}[t]
\caption{
Adaptive Sampling Process}\label{alg:adp}
%\begin{small}
%\begin{scriptsize}
\begin{flushleft}
        \textbf{INPUT:} 
        Recommender Model $RS$, test user set $\mathcal{U}$, initial size $n_0$, terminal size $n_{max}$
        \\
        \textbf{OUTPUT:} $\{(u, r_u, n_u)\}$
\end{flushleft}
\vspace{-1em}
\begin{algorithmic}[1]
% \STATE $X^TX= V \Sigma^T \Sigma V^T $ \text{ (Eigen Decomposition) }
\FORALL{$u \in \mathcal{U}$} 
    \STATE sampling $n_0 - 1$ items, form the sample set $I^s_u$
    
    \STATE $n_u = n_0$, $r_u = RS(i_u, I^s_u)$
    
    \WHILE{$r_u = 1$ and $n_u \neq n_{max}$}
    \STATE ${}$\hspace{2em} sampling extra $n_u$ items, form the new set $I^s_u$
    \STATE ${}$\hspace{2em} $n_u = 2n_u$, $r_u = RS(i_u, I^s_u)$
    \ENDWHILE
    \STATE record $n_u, r_u$ for user $u$
\ENDFOR
\end{algorithmic}
%\end{small}
%\end{scriptsize}
\end{algorithm}

% Next, we will discuss how to use the adaptive item sampling to estimate the global top-$K$ metrics. 
\subsection{Maximum Likelihood Estimation by EM}\label{MLE}

To utilize the adaptive item sampling for estimating the global top-$K$ metrics, we review two routes: 1) approaches from ~\cite{Krichene20@KDD20} and our aforementioned new estimators in this paper; 2)  methods based on MLE and EM ~\cite{Jin@AAAI21}. Since every user has a different number of item samples, we found the first route is hard to extend (which requires an equal sample size); but luckily the second route is much more flexible and can be easily generalized to this situation. 

To begin with, we note that for any user $u$ (his/her test item ranks $r_u$ in the sample set (with size $n_u$) and ranks $R_u$ (unknown)), its rank $r_u$ follows a binomial distribution: 
\begin{equation}
\begin{split}
    P(r=r_u|R=R_u; n_u)=Bin(r_u -1;n_u -1,\theta_u)\\
    %&=\binom{n_u-1}{r_u-1}\theta_u^{r_u-1}(1-\theta_u)^{n_u-r_u}
    \end{split}
\end{equation}

Given this, let $\pmb{\Pi} = (\pi_1,\dots,\pi_R,\dots,\pi_N)^T$ be the parameters of the mixture of binomial distributions, $\pi_R$ is the probability for user population ranks at position $R$ globally. And then we have $p(r_u|\pmb{\Pi})=\sum_{R=1}^N \pi_R \cdot p(r_u|\theta_R; n_u)$, where 
$p(r_u|\theta_R; n_u)=Bin(r_u-1; n_u-1, \theta_R)$. 
We can apply the maximal likelihood estimation (MLE) to learn the parameters of the mixture of binomial distributions ($MB$), which naturally generalizes the EM procedure (details see \cref{sec:app_em}) used in ~\cite{Jin@AAAI21}, where each user has the same $n$ samples: 

\begin{equation*}
    \begin{split}
        \phi(R_{uk}) &= P(R_u = k|r_u;\boldsymbol{\pi}^{old})\\
\pi^{new}_k&=\frac{1}{M}\sum\limits_{u=1}^M\phi(R_{uk})   
    \end{split}
\end{equation*}

When the process converges, we obtain $\pmb{\Pi}^*$ and use it to estimate $\pmb{P}$, i.e., $\widehat{P}(R)=\pi^*_R$. 
Then, we can use $\widehat{P}(R)$ to estimate the desired metric $metric@K$. 
The overall time complexity is linear with respect to the sample size $O(t\sum{n_u})$ where $t$ is the iteration number. 

\subsection{Sampling Size UpperBound}
\label{upperbound}

\begin{figure}
    \centering
    \includegraphics[ width=0.72\linewidth]{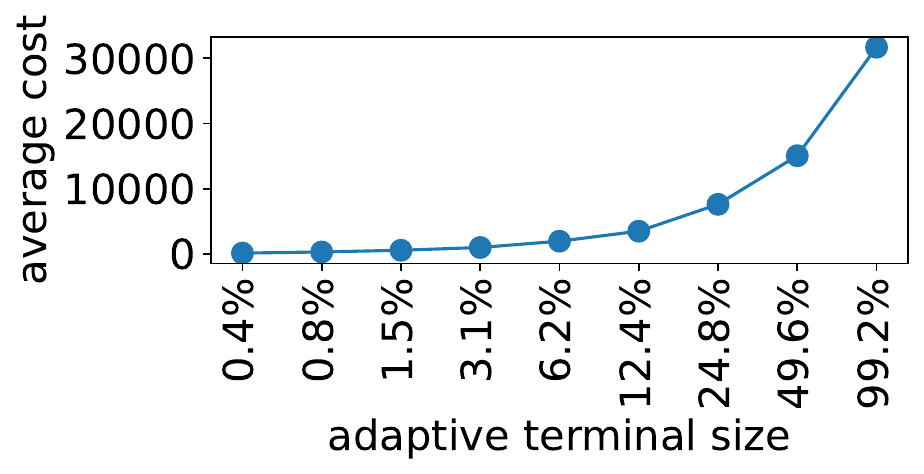}
    \caption{Sample efficiency w.r.t terminal size (ratio of total items). The illustration result is obtained by EASE model \cite{Steck_2019} over $yelp$ dataset.}
    \label{fig:sample_efficiency}
\end{figure}

Now, we consider how to determine the terminal size $n_{max}$. We take the post-analysis over the different terminal sizes and investigate the average sampling cost, which introduces the concept \textit{sampling efficiency}, see \cref{fig:sample_efficiency}. Formally, we first select a large number $n_{max} \approx N$ and repeat the aforementioned adaptive sampling process. For each user, his/her sampling set size could be one of $\{n_0, n_1=2n_0, n_2=4n_0,\dots, n_t=n_{max}\}$. And there are $m_j$ users whose sample set size is $n_j$ $(j=0,1,\dots, t)$.
The average sampling cost for each size $n_j$ can be defined heuristically:
\begin{equation}\label{eq:ave_cost}
    \begin{split}
        C_j &= \frac{ (M-\sum\limits_{p=0}^{j-1}{m_p}) \times (n_j-n_{j-1}) }{m_j}\quad j\neq 0, t\\
        C_0 &= \frac{M\times n_0}{m_0}
    \end{split}
\end{equation}
The intuition behind \cref{eq:ave_cost} is: at $j$-th iteration, we independently sample $n_j-n_{j-1}$ items for total $M-\sum\limits_{p=0}^{j-1}{m_p}$ users, and there are $m_j$ users whose rank $r_u> 1$. $C_j$ is the average items to be sampled to get a user whose $r_u> 1$, which reflects sampling efficiency. In ~\cref{fig:sample_efficiency}, we can see that when the sample reaches $12.4\%$ (of total items, around $3200$ for $yelp$ dataset) the sampling efficiency will reduce quickly (the average cost $C_j$ increases fast). Such post-analysis provides insights into how to balance the sample size and sampling efficiency. In this case, we observe $12.4\%$ can be a reasonable choice. Even though different datasets can pick up different thresholds, we found in practice $10\%\sim 15\%$  can serve as a default choice to start and achieve pretty good performance for the estimation accuracy.

\section{Experiments}\label{sec:exp}

\begin{table*}[]
\centering
% \usepackage{multirow}
% \usepackage{graphicx}
\resizebox{0.9\textwidth}{!}{%
\begin{tabular}{|c|c|rrrrrrr|}
\hline
\multirow{3}{*}{dataset} &
  \multirow{3}{*}{Models} &
  \multicolumn{7}{c|}{sample set size 100} \\ \cline{3-9} 
 &
   &
  \multicolumn{3}{c|}{baseline} &
  \multicolumn{4}{c|}{this paper} \\ \cline{3-9} 
 &
   &
  \multicolumn{1}{c|}{MES} &
  \multicolumn{1}{c|}{MLE} &
  \multicolumn{1}{c|}{BV} &
  \multicolumn{1}{c|}{BV\_MES} &
  \multicolumn{1}{c|}{BV\_MLE} &
  \multicolumn{1}{c|}{MN\_MES} &
  \multicolumn{1}{c|}{MN\_MLE} \\ \hline
\multirow{5}{*}{pinterest-20} &
  EASE &
  \multicolumn{1}{r|}{5.86$\rpm$2.26} &
  \multicolumn{1}{r|}{5.54$\rpm$1.85} &
  \multicolumn{1}{r|}{8.11$\rpm$2.00} &
  \multicolumn{1}{r|}{\textbf{5.05$\rpm$1.46}} &
  \multicolumn{1}{r|}{5.14$\rpm$1.46} &
  \multicolumn{1}{r|}{\textbf{5.00$\rpm$1.39}} &
  5.10$\rpm$1.34 \\ \cline{2-9} 
 &
  MultiVAE &
  \multicolumn{1}{r|}{4.17$\rpm$2.91} &
  \multicolumn{1}{r|}{3.34$\rpm$2.07} &
  \multicolumn{1}{r|}{\textbf{2.75$\rpm$1.61}} &
  \multicolumn{1}{r|}{2.89$\rpm$1.74} &
  \multicolumn{1}{r|}{\textbf{2.88$\rpm$1.74}} &
  \multicolumn{1}{r|}{\textbf{2.75$\rpm$1.66}} &
  \textbf{2.75$\rpm$1.68} \\ \cline{2-9} 
 &
  NeuMF &
  \multicolumn{1}{r|}{5.17$\rpm$2.74} &
  \multicolumn{1}{r|}{4.28$\rpm$1.95} &
  \multicolumn{1}{r|}{4.23$\rpm$1.79} &
  \multicolumn{1}{r|}{3.83$\rpm$1.59} &
  \multicolumn{1}{r|}{3.84$\rpm$1.72} &
  \multicolumn{1}{r|}{\textbf{3.60$\rpm$1.50}} &
  \textbf{3.76$\rpm$1.44} \\ \cline{2-9} 
 &
  itemKNN &
  \multicolumn{1}{r|}{5.90$\rpm$2.20} &
  \multicolumn{1}{r|}{5.80$\rpm$1.60} &
  \multicolumn{1}{r|}{8.93$\rpm$1.70} &
  \multicolumn{1}{r|}{\textbf{5.11$\rpm$1.22}} &
  \multicolumn{1}{r|}{5.31$\rpm$1.25} &
  \multicolumn{1}{r|}{\textbf{5.09$\rpm$1.15}} &
  5.26$\rpm$1.14 \\ \cline{2-9} 
 &
  ALS &
  \multicolumn{1}{r|}{4.19$\rpm$2.37} &
  \multicolumn{1}{r|}{3.44$\rpm$1.68} &
  \multicolumn{1}{r|}{3.17$\rpm$1.34} &
  \multicolumn{1}{r|}{3.05$\rpm$1.39} &
  \multicolumn{1}{r|}{3.07$\rpm$1.42} &
  \multicolumn{1}{r|}{\textbf{2.86$\rpm$1.27}} &
  \textbf{2.90$\rpm$1.28} \\ \hline
\multirow{5}{*}{yelp} &
  EASE &
  \multicolumn{1}{r|}{8.08$\rpm$4.94} &
  \multicolumn{1}{r|}{7.89$\rpm$4.70} &
  \multicolumn{1}{r|}{18.60$\rpm$2.78} &
  \multicolumn{1}{r|}{6.10$\rpm$3.74} &
  \multicolumn{1}{r|}{6.56$\rpm$3.90} &
  \multicolumn{1}{r|}{\textbf{4.84$\rpm$2.17}} &
  \textbf{5.61$\rpm$2.30} \\ \cline{2-9} 
 &
  MultiVAE &
  \multicolumn{1}{r|}{9.33$\rpm$6.61} &
  \multicolumn{1}{r|}{7.67$\rpm$4.94} &
  \multicolumn{1}{r|}{9.70$\rpm$3.22} &
  \multicolumn{1}{r|}{6.84$\rpm$4.10} &
  \multicolumn{1}{r|}{6.80$\rpm$4.04} &
  \multicolumn{1}{r|}{\textbf{4.30$\rpm$1.27}} &
  \textbf{4.35$\rpm$1.31} \\ \cline{2-9} 
 &
  NeuMF &
  \multicolumn{1}{r|}{15.09$\rpm$6.24} &
  \multicolumn{1}{r|}{15.47$\rpm$5.55} &
  \multicolumn{1}{r|}{22.40$\rpm$3.17} &
  \multicolumn{1}{r|}{\textbf{13.14$\rpm$4.55}} &
  \multicolumn{1}{r|}{13.92$\rpm$4.70} &
  \multicolumn{1}{r|}{\textbf{13.46$\rpm$2.43}} &
  14.50$\rpm$2.45 \\ \cline{2-9} 
 &
  itemKNN &
  \multicolumn{1}{r|}{9.25$\rpm$4.87} &
  \multicolumn{1}{r|}{9.62$\rpm$4.88} &
  \multicolumn{1}{r|}{23.24$\rpm$2.16} &
  \multicolumn{1}{r|}{\textbf{7.69$\rpm$4.09}} &
  \multicolumn{1}{r|}{8.15$\rpm$4.17} &
  \multicolumn{1}{r|}{\textbf{7.74$\rpm$2.08}} &
  8.75$\rpm$2.08 \\ \cline{2-9} 
 &
  ALS &
  \multicolumn{1}{r|}{14.31$\rpm$3.96} &
  \multicolumn{1}{r|}{13.68$\rpm$3.51} &
  \multicolumn{1}{r|}{15.14$\rpm$1.86} &
  \multicolumn{1}{r|}{13.43$\rpm$3.16} &
  \multicolumn{1}{r|}{13.26$\rpm$3.08} &
  \multicolumn{1}{r|}{\textbf{11.68$\rpm$0.88}} &
  \textbf{11.57$\rpm$0.83} \\ \hline
\end{tabular}%
}
\caption{The average relative errors between estimated $Recall@K$ ($K$ from $1$ to $50$) and the true ones. Unit is $\%$. In each row, the smallest two results are highlighted in bold, indicating the most accurate results. Sample set size $n=100$.}
\label{tab:recall_error_100}
% \end{small}
\end{table*}

In this section, we evaluate the performance of the new proposed estimator in \cref{eq:mn_closed} compared to the baselines from \cite{Krichene20@KDD20,Jin@AAAI21} and validate the effectiveness and efficiency of adaptive sampling-based MLE (\textbf{adaptive MLE}).
Specifically, we aim to answer three questions: 

\noindent{\textbf{Question 1.}} How do the new estimators in ~\cref{sec:newestimator} perform compared to estimators based on learning empirical distributions (i.e., $MLE$, $MES$ in~\cite{Jin@AAAI21}), and to the $BV$ approach in~\cite{Krichene20@KDD20}?

\noindent{\textbf{Question 2.}} 
How effective and efficient the adaptive item-sampling evaluation method \textbf{adaptive MLE} is, compared with the best estimators for the basic (non-adaptive) item sampling methods in ~\cref{sec:newestimator}? 

\noindent{\textbf{Question 3.}} 
How accurately can these estimators find the best model (in terms of the global top-K metric) among a list of recommendation models?

\subsubsection{Experimental Setup}
We take three widely-used datasets for recommendation system research, $pinterest-20$, $yelp$, $ml-20m$. See also \cref{sec:app_exp} for dataset statistics. We follow the work~\cite{Jin@AAAI21} to adopt five popular recommendation algorithms, including three non-deep-learning methods (itemKNN~\cite{DeshpandeK@itemKNN}, ALS~\cite{hu2008collaborative}, and EASE~\cite{Steck_2019}) 
and two deep learning ones (NeuMF~\cite{he2017neural} and MultiVAE~\cite{liang2018variational}).
Three most popular top-K metrics: $Recall$, $NDCG$, and $AP$ are utilized for evaluating the recommendation models. 

\subsubsection{Estimating Procedure} There are $M$ users and $N$ items. 
Each user $u$ is associated with a target item $i_u$. The learned recommendation algorithm/model $A$ would compute the ranks $\{R_u\}_{u=1}^M$ among all items called global ranks and the ranks $\{r_u\}_{u=1}^M$ among sampled items called sampled ranks. Without the knowledge of $\{R_u\}_{u=1}^M$, the estimator tries to estimate the global metric defined in \cref{eq:topk_metric} based on $\{r_u\}_{u=1}^M$. We repeat experiments $100$ times, deriving $100$ distinct $\{r_u\}_{u=1}^M$ set. Below reports the experimental results that are averaged on these {$\mathbf{100}$} repeat.

Due to the space limitation, we report representative experimental results in the following and leave baselines, dataset statistics, and more results in the \cref{sec:app_exp}.

\subsection*{Q1. Accuracy of the Estimators for Basic Item-Sampling}
Here, we aim to answer Question 1: comparing the new estimators in ~\cref{sec:newestimator} perform against the state-of-the-art methods from ~\cite{Krichene20@KDD20, Jin@AAAI21} under the basic item sampling scheme. 
 Different from \cite{Jin@AAAI21}, where authors list all the estimating as well as the true results for $metric@10$ and conclude, here we would quantify the accuracy of each estimator in terms of relative error, leading to a more rigorous and reliable comparison. Specifically, we compute the true global $metric@k$ ( $k$ from $1$ to $50$), then we average the absolute relative error between the estimated $metric@k$ from each estimator and the true one.
 
 The estimators include $BV$ (with the tradeoff parameter $\gamma=0.01$) from \cite{Krichene20@KDD20}, $MLE$ (Maximal Likelihood Estimation), $MES$ (Maximal Entropy with Squared distribution distance, where $\eta=0.001$) from \cite{Jin@AAAI21} and newly proposed estimators \cref{eq:mn_closed}. \Cref{tab:recall_error_100} (see Appendix for a complete result version.) presents the average relative error of the estimators in terms of $Recall@K$ ($k$ from $1$ to $50$). The results of $NDCG@K$ and $AP@K$ are in \cref{sec:app_exp}. We highlight the most and the second-most accurate estimator. For instance, for model $EASE$ in dataset $pinterest-20$ (line $1$ of \Cref{tab:recall_error_100}), the estimator $MN\_MES$ is the most accurate one with $5.00\%$ average relative error compared to its global $Recall@K$ ($K$ from $1$ to $50$).
 
Overall, we observe from \Cref{tab:recall_error_100} that $MN\_MES$ and $MN\_MLE$ are among the most or the second-most accurate estimators.  And in most cases, they outperform the others significantly. Meantime, they have a smaller deviation compared to their prior estimators $MES$ and $MLE$. In addition, we also notice that the estimators with the knowledge of some reasonable prior distribution ($BV\_MES$, $MN\_MES$, $BV\_MLE$, $MN\_MLE$) could achieve more accurate results than the others. This indicates that these estimators could better help the distribution to converge.

\subsection*{Q2. Performance of Adaptive Sampling}
Here, we aim to answer Question 2: comparing the adaptive item sampling method \textbf{adaptive MLE}, comparing with the best estimators for the basic (non-adaptive) item sampling methods, such as $BV\_MES$, $MN\_MES$, $BV\_MLE$, $MN\_MLE$.

\begin{table*}[]
\centering
% \usepackage{multirow}
% \usepackage{graphicx}
\resizebox{0.9\textwidth}{!}{%
\begin{tabular}{|cc|rrrr|rr|}
\hline
\multicolumn{1}{|c|}{\multirow{2}{*}{Dataset}} &
  \multirow{2}{*}{Models} &
  \multicolumn{4}{c|}{Fix Sample} &
  \multicolumn{2}{c|}{Adaptive Sample} \\ \cline{3-8} 
\multicolumn{1}{|c|}{} &
   &
  \multicolumn{1}{c|}{BV\_MES} &
  \multicolumn{1}{c|}{BV\_MLE} &
  \multicolumn{1}{c|}{MN\_MES} &
  \multicolumn{1}{c|}{MN\_MLE} &
  \multicolumn{1}{c|}{average size} &
  \multicolumn{1}{c|}{adaptive MLE} \\ \hline
\multicolumn{2}{|c|}{} &
  \multicolumn{4}{c|}{sample set size n = 500} &
  \multicolumn{2}{c|}{} \\ \hline
\multicolumn{1}{|c|}{\multirow{5}{*}{pinterest-20}} &
  EASE &
  \multicolumn{1}{r|}{3.87$\rpm$2.13} &
  \multicolumn{1}{r|}{4.33$\rpm$2.23} &
  \multicolumn{1}{r|}{4.17$\rpm$2.45} &
  4.33$\rpm$2.50 &
  \multicolumn{1}{r|}{307.74$\rpm$1.41} &
  \textbf{1.46$\rpm$0.63} \\ \cline{2-8} 
\multicolumn{1}{|c|}{} &
  MultiVAE &
  \multicolumn{1}{r|}{2.66$\rpm$1.75} &
  \multicolumn{1}{r|}{2.58$\rpm$1.75} &
  \multicolumn{1}{r|}{3.26$\rpm$2.14} &
  3.07$\rpm$2.09 &
  \multicolumn{1}{r|}{286.46$\rpm$1.48} &
  \textbf{1.67$\rpm$0.70} \\ \cline{2-8} 
\multicolumn{1}{|c|}{} &
  NeuMF &
  \multicolumn{1}{r|}{2.82$\rpm$1.98} &
  \multicolumn{1}{r|}{2.79$\rpm$1.96} &
  \multicolumn{1}{r|}{3.24$\rpm$2.30} &
  3.22$\rpm$2.27 &
  \multicolumn{1}{r|}{259.77$\rpm$1.28} &
  \textbf{1.73$\rpm$0.83} \\ \cline{2-8} 
\multicolumn{1}{|c|}{} &
  itemKNN &
  \multicolumn{1}{r|}{3.99$\rpm$2.22} &
  \multicolumn{1}{r|}{4.18$\rpm$2.24} &
  \multicolumn{1}{r|}{4.23$\rpm$2.47} &
  4.06$\rpm$2.44 &
  \multicolumn{1}{r|}{309.56$\rpm$1.31} &
  \textbf{1.42$\rpm$0.67} \\ \cline{2-8} 
\multicolumn{1}{|c|}{} &
  ALS &
  \multicolumn{1}{r|}{3.35$\rpm$1.89} &
  \multicolumn{1}{r|}{3.24$\rpm$1.87} &
  \multicolumn{1}{r|}{3.90$\rpm$2.24} &
  3.80$\rpm$2.23 &
  \multicolumn{1}{r|}{270.75$\rpm$1.22} &
  \textbf{1.84$\rpm$1.07} \\ \hline
\multicolumn{2}{|c|}{} &
  \multicolumn{4}{c|}{sample set size n = 500} &
  \multicolumn{2}{c|}{} \\ \hline
\multicolumn{1}{|c|}{\multirow{5}{*}{yelp}} &
  EASE &
  \multicolumn{1}{r|}{5.63$\rpm$3.73} &
  \multicolumn{1}{r|}{5.48$\rpm$3.66} &
  \multicolumn{1}{r|}{4.03$\rpm$2.53} &
  4.04$\rpm$2.53 &
  \multicolumn{1}{r|}{340.79$\rpm$2.03} &
  \textbf{3.55$\rpm$2.00} \\ \cline{2-8} 
\multicolumn{1}{|c|}{} &
  MultiVAE &
  \multicolumn{1}{r|}{7.68$\rpm$5.86} &
  \multicolumn{1}{r|}{7.48$\rpm$5.74} &
  \multicolumn{1}{r|}{5.77$\rpm$3.87} &
  5.64$\rpm$3.82 &
  \multicolumn{1}{r|}{288.70$\rpm$2.24} &
  \textbf{5.09$\rpm$2.60} \\ \cline{2-8} 
\multicolumn{1}{|c|}{} &
  NeuMF &
  \multicolumn{1}{r|}{9.34$\rpm$5.50} &
  \multicolumn{1}{r|}{9.55$\rpm$5.50} &
  \multicolumn{1}{r|}{8.43$\rpm$4.07} &
  8.91$\rpm$4.13 &
  \multicolumn{1}{r|}{290.62$\rpm$2.11} &
  \textbf{4.43$\rpm$2.55} \\ \cline{2-8} 
\multicolumn{1}{|c|}{} &
  itemKNN &
  \multicolumn{1}{r|}{5.01$\rpm$2.99} &
  \multicolumn{1}{r|}{4.99$\rpm$2.96} &
  \multicolumn{1}{r|}{\textbf{3.65$\rpm$2.28}} &
  3.71$\rpm$2.29 &
  \multicolumn{1}{r|}{369.16$\rpm$2.51} &
  3.67$\rpm$2.73 \\ \cline{2-8} 
\multicolumn{1}{|c|}{} &
  ALS &
  \multicolumn{1}{r|}{13.39$\rpm$7.34} &
  \multicolumn{1}{r|}{13.94$\rpm$7.61} &
  \multicolumn{1}{r|}{12.57$\rpm$5.46} &
  13.67$\rpm$5.80 &
  \multicolumn{1}{r|}{297.07$\rpm$2.29} &
  \textbf{5.48$\rpm$3.34} \\ \hline
\end{tabular}%
}
\caption{The average relative errors between estimated $NDCG@K$ ($K$ from $1$ to $50$) and the true ones. Unit is $\%$. In each row, the smallest relative error is highlighted, indicating the most accurate result.}
\label{tab:ndcg_adaptive}
\end{table*}

\begin{table*}[]
\centering
\resizebox{0.9\textwidth}{!}{%
\begin{tabular}{|ccc|cccc|c|}
\hline
\multicolumn{1}{|c|}{\multirow{3}{*}{Dataset}} &
  \multicolumn{1}{c|}{\multirow{3}{*}{Top-K}} &
  \multirow{3}{*}{Metrics} &
  \multicolumn{4}{c|}{Fix Sample} &
  AdaptiveSample \\ \cline{4-8} 
\multicolumn{1}{|c|}{} &
  \multicolumn{1}{c|}{} &
   &
  \multicolumn{1}{c|}{BV\_MES} &
  \multicolumn{1}{c|}{BV\_MLE} &
  \multicolumn{1}{c|}{MN\_MES} &
  MN\_MLE &
  adaptive MLE \\ \cline{4-8} 
\multicolumn{1}{|c|}{} &
  \multicolumn{1}{c|}{} &
   &
  \multicolumn{4}{c|}{sample set size n = 500} &
  sample set size $260\sim 310$ \\ \hline
\multicolumn{1}{|c|}{\multirow{6}{*}{pinterest-20}} &
  \multicolumn{1}{c|}{\multirow{3}{*}{10}} &
  RECALL &
  \multicolumn{1}{c|}{69} &
  \multicolumn{1}{c|}{73} &
  \multicolumn{1}{c|}{67} &
  69 &
  78 \\ \cline{3-8} 
\multicolumn{1}{|c|}{} &
  \multicolumn{1}{c|}{} &
  NDCG &
  \multicolumn{1}{c|}{58} &
  \multicolumn{1}{c|}{59} &
  \multicolumn{1}{c|}{58} &
  60 &
  84 \\ \cline{3-8} 
\multicolumn{1}{|c|}{} &
  \multicolumn{1}{c|}{} &
  AP &
  \multicolumn{1}{c|}{54} &
  \multicolumn{1}{c|}{57} &
  \multicolumn{1}{c|}{52} &
  52 &
  68 \\ \cline{2-8} 
\multicolumn{1}{|c|}{} &
  \multicolumn{1}{c|}{\multirow{3}{*}{20}} &
  RECALL &
  \multicolumn{1}{c|}{69} &
  \multicolumn{1}{c|}{73} &
  \multicolumn{1}{c|}{70} &
  74 &
  81 \\ \cline{3-8} 
\multicolumn{1}{|c|}{} &
  \multicolumn{1}{c|}{} &
  NDCG &
  \multicolumn{1}{c|}{69} &
  \multicolumn{1}{c|}{73} &
  \multicolumn{1}{c|}{68} &
  73 &
  79 \\ \cline{3-8} 
\multicolumn{1}{|c|}{} &
  \multicolumn{1}{c|}{} &
  AP &
  \multicolumn{1}{c|}{57} &
  \multicolumn{1}{c|}{60} &
  \multicolumn{1}{c|}{54} &
  56 &
  69 \\ \hline
\end{tabular}%
}
\caption{Accuracy of predicting the winner models for different datasets. Values in the table are the number of correct predictions over 100 repeats. The larger number, the better estimator}
\label{tab:winner}
\end{table*}

 \Cref{tab:ndcg_adaptive} (complete version in appendix ) presents the average relative error of the estimators in terms of $NDCG@K$ ($k$ from $1$ to $50$).  We highlight the most accurate estimator. For the basic item sampling, we choose $500$ sample size for datasets $pinterest-20$ and $yelp$, and $1000$ sample size for dataset $ml-20m$. The upperbound threshold $n_{max}$ is set to $3200$. 
 
 We observe that adaptive sampling uses much less sample size (typically $200-300$ vs $500$ on $pinterest-20$, $yelp$ datasets and $700-800$ vs $1000$ on $ml-20m$ dataset). Particularly, the relative error of the adaptive sampling is significantly smaller than that of the basic sampling methods. On the first ($pinterest-20$) and third ($ml-20m$) datasets, the relative errors have reduced to less than $2\%$. In other words, the adaptive method has been much more effective (in terms of accuracy) and efficient (in terms of sample size).  This also confirms the benefits of addressing the ``blind spot'' issue, which provides higher resolution to recover global $K$ metrics for small $K$ ($K\leq 50$ here).

\subsection*{Q3. Estimating the Winner}

\Cref{tab:winner} (complete version in appendix) indicates the results of among the $100$ repeats, how many times that an estimator could match the best recommendation algorithms for a given $metric@K$. More concretely, for a global $metric@K$ (compute from \cref{eq:topk_metric}), there exists a recommendation algorithm/model which performs best called the winner. The estimator could also estimate $metric@K$ for each algorithm based on its $\{r_u\}_{u=1}^M$. Among this estimated $ metric@K$, one can also find the best recommendation model. Intuitively, if the estimator is accurate enough, it could find the same winner as the truth. Thus, we count the success time that an estimator can find as another measure of estimating accuracy. From the \Cref{tab:winner}, We observe that new proposed adaptive estimators could achieve competitive or even better results to the baselines with much less average sample cost.

\section{Conclusion}
\label{sec:conc}
In this paper, we propose first item-sampling estimators which explicitly optimize its mean square error with respect to the ground truth. Then we highlight the subtle difference between the estimators from \cite{Krichene20@KDD20} and ours, and point out the potential issue of the former - failing to link the user size with the variance. Furthermore, we address the limitation of the current item sampling approach, which typically does not have sufficient granularity to recover the top $K$ global metrics especially when $K$ is small. We then propose an effective adaptive item-sampling method. The experimental evaluation demonstrates the new adaptive item sampling significantly improves both the sampling efficiency and estimation accuracy. Our results provide a solid step toward making item sampling available for recommendation research and practice. In the future, we would like to further investigate how to combine item-sampling working with user sampling to speed up the offline evaluation.

\section{Acknowledgement}

This work was partially supported by the National Science Foundation under Grant IIS-2142675.

\bibliography{ref}

\appendix
\newpage
\section{Related Work: Efforts in Metric Estimation}
\label{sec:app_related}
There are two recent works studying the general metric estimation problem based on item sampling metrics. Specifically, given the sampling ranked results in the test set, $\{r_u\}^{M}_{u=1}$, how to infer/approximate the $T$ from \cref{eq:metric_0} or more commonly \cref{eq:topk_metric}, without the knowledge $\{R_u\}_{u=1}^M$?

\noindent{\bf Krichene and Rendle's approaches:} ~\cite{Krichene20@KDD20} develop a discrete function $\widehat{\mathcal M}(r)$ so that: 
\begin{small}
\begin{equation}
\begin{split}
    T&=\frac{1}{M}\sum_{u=1}^M \mathbf{1}_{R_u\le K}\cdot {\mathcal M}(R_u) \\
    &\approx \frac{1}{M}\sum_{u=1}^M \widehat{\mathcal M}(r_u) = \sum_{r=1}^n \tilde{P}(r) \widehat{\mathcal M}(r)=\widehat{T}
    \end{split}
\end{equation}
\end{small}
where $\tilde{P}(r)$ is the empirical rank distribution on the sampling data (~\cref{tab:notations}). They have proposed a few estimators based on this idea, including estimators that use the unbiased rank estimators,  minimize bias with monotonicity constraint ($CLS$), and utilize Bias-Variance ($BV$) tradeoff. Their study shows that only the last one ($BV$) is competitive. They further derive a solution based on the Bias-Variance ($BV$) tradeoff:
\begin{equation} \label{eq:leastsquare}
 \begin{split}
&\widehat{\mathcal{M}} = \Big((1.0-\gamma)A^TA+\gamma \text{diag}(\pmb{c})  \Big)^{-1}A^T\pmb{b}\\
        &A\in \mathbb{R}^{N\times n},\quad A_{R,r} = \sqrt{P(R)}P(r|R)\\
        &\pmb{b}\in \mathbb{R}^N,\quad b_{R} = \sqrt{P(R)}\mathcal{M}(R)\\
        &\pmb{c}\in \mathbb{R}^n, \quad c_{r} = \sum\limits_R^{N}{P(R)P(r|R)}
    \end{split}
\end{equation}

\noindent{\bf Jin et al.'s Method:} 
\cite{Jin@AAAI21} proposed different types of estimators based on learning the empirical rank distribution $(P(R))_{R=1}^N$:
\begin{small}
\begin{equation}\label{eq:metric_1}
T=\frac{1}{M}\sum_{u=1}^M \mathbf{1}_{R_u\le K}\cdot {\mathcal M}(R_u)=\sum_{R=1}^K P(R) \cdot {\mathcal M}(R) 
\end{equation}
\end{small}
Thus, if $P(R)$ can be estimated (by $\widehat{P}(R)$), then the metric \cref{eq:metric_1} can be estimated by:
\begin{equation}\label{eq:metric_ap}
\widehat{T}=\sum_{R=1}^K \widehat{P}(R) \cdot  {\mathcal M}(R)
\end{equation}
They further proposed two approaches to learn the empirical rank distribution $(P(R))_{R=1}^N$. The first approach is to learn the parameters of the mixture of binomial distributions ($MB$) given $\{r_u\}_{u=1}^M$ based on maximal likelihood estimation (MLE). The other approach for estimating a (discrete) probability distribution is based on the principal of maximal entropy~\cite{elementsinfo}. 

\section{Notation}
\label{sec:note}

\begin{table}
 
      %\begin{minipage}{0.48\textwidth}
      %\vspace{-2.0ex}
 \begin{tabular}{|l|l|}
 %\hline
 %\noindent
\hline
$M$ & \# of users in {\bf test} set \\
\hline
$N$ & \# of total items \\
\hline
$I$	& the set of all items, and $|I| = N$ \\
\hline
$i_u$	& target item for user $u$ in {\bf test} set\\
\hline 
$R_u$	& rank of item $i_u$ among $I$ for user $u$	\\
\hline 
$n$ & sample set size (1 target item, $n-1$ sampled items) \\
\hline
$I_u$	& $I_u \backslash i_u$ consists of $n-1$ sampled items for $u$ \\
\hline
$r_u$	& rank of item $i_u$ among $I_u$\\
\hline
$T$ & evaluation metric, e.g. $Recall$ \\
\hline
$T_S$ & sampled evaluation metric, e.g. $Recall_S$ \\
\hline
$\widehat{T}$ & estimated evaluation metric\\
\hline
  \end{tabular}
   \caption{Definition of notations}
   \label{tab:notations}
  %\end{minipage}
  \end{table}
  
\section{Rewriting of \cref{eq:last0}}\label{sec:mse_re}

\begin{small}
\begin{equation}\label{eqn:last_1}
    \begin{split}
        &\E[\widehat{T}-\sum_{R=1}^N P(R) \mathcal{M}(R)]^2 \\
 =&\E[\E \widehat{T}-\sum_{R=1}^N P(R) \mathcal{M}(R) + \widehat{T} - \E \widehat{T}]^2\nonumber \\
 =&\Big(\E \widehat{T}- \sum_{R=1}^N P(R) \mathcal{M}(R)\Big)^2 + \E[\widehat{T}-\E \widehat{T}]^2 \nonumber \\
=&\Big(\sum_{r=1}^n P(r) \widehat{\mathcal M}(r)-\sum_{R=1}^N P(R) \mathcal{M}(R)\Big)^2  \\
 +&\E[\sum_{r=1}^n \tilde{P}(r) \widehat{\mathcal M}(r)-\sum_{r=1}^n P(r) \widehat{\mathcal M}(r)]^2 \nonumber \\
  =&  \Big(\sum_{r=1}^n \sum_{R=1}^N P(r|R)P(R) \widehat{\mathcal M}(r)-\sum_{R=1}^N P(R) \mathcal{M}(R)\Big)^2  \nonumber \\+& \E[\sum_{r=1}^n \sum_{R=1}^N \tilde{P}(r|R)P(R) \widehat{\mathcal M}(r)- \sum_{r=1}^n \sum_{R=1}^N  P(r|R)P(R) \widehat{\mathcal M}(r)]^2 
    \end{split}
\end{equation}
\end{small}

\section{Linear Combination of Coefficients}\label{sec:app_linear}
Considering $X=(X_1,\dots,X_n)$ is the random variables of a sample $M$ times multinomial distribution with $n$ cells probabilities $(\theta_1,\dots,\theta_n)$. We have:$\frac{X_i}{M}\rightarrow \theta_i, \text{ when } M\rightarrow \infty$
\begin{equation}
    \begin{split}
        &\E[{X_i}] = M\theta_i\quad Var[X_i] = M\theta_i(1-\theta_i)\\
        &Cov(X_i, X_j) = -M\theta_i\theta_j
    \end{split}
\end{equation}
Considering the a new random variable deriving from the linear combination: $\mathcal{A} = \sum\limits_{i=1}^n{w_iX_i}$, where the ${w_i}$ are the constant coefficients.
\begin{small}
\begin{equation*}\label{eq:A_equation}
    \begin{split}
    \E[\mathcal{A}]&=  M\cdot\sum\limits_{i=1}^n{w_i\theta_i}\\
        Var[\mathcal{A}] &= \E[\mathcal{A}^2]-(\E[\mathcal{A}])^2=\sum\limits_i^{n}{w^2_i[M\theta_i-M\theta_i^2]}-2\sum\limits_{i\neq j}{w_iw_j[M\theta_i\theta_j]}\\
        &=M\sum\limits_{i}^n{w^2_i\theta_i}-M\cdot\Big(\sum\limits_{i}^n{w^2_i\theta^2_i}+2\sum\limits_{i\neq j}w_iw_j\theta_i\theta_j\Big)\\
        &=M\cdot\Big(\sum\limits_{i}^n{w^2_i\theta_i}-\big(\sum\limits_{i}^n{w_i\theta_i}\big)^2 \Big)
    \end{split}
\end{equation*}
\end{small}

\section{Rewriting of $L_2$}\label{sec:l2_re}

\begin{equation}
    \begin{split}
    \mathcal{L}_2&=\sum\limits_{R=1}^N{P(R)\cdot \E[\sum_{r=1}^n \tilde{P}(r|R) \widehat{\mathcal M}(r)- \sum_{r=1}^n  P(r|R) \widehat{\mathcal M}(r)]^2 }\\
        &=\sum\limits_{R=1}^N{P(R)\cdot \E[\sum_{r=1}^n \frac{X_r}{M\cdot P(R)} \widehat{\mathcal M}(r)- \sum_{r=1}^n  \frac{\E[X_r]}{M\cdot P(R)} \widehat{\mathcal M}(r)]^2 }\\
        &=\sum\limits_{R=1}^N{\frac{1}{M^2\cdot P(R)}\cdot \E[\sum_{r=1}^n {X_r}\widehat{\mathcal M}(r)- \sum_{r=1}^n  {\E[X_r]} \widehat{\mathcal M}(r)]^2 }\\
 &= \sum\limits_{R=1}^{N}{\frac{1}{M}\cdot \Big(\sum\limits_{r}^n{\widehat{\mathcal M}^2(r) P(r|R)}-\big(\sum\limits_{r}^n{\widehat{\mathcal M}(r)P(r|R)}\big)^2 \Big) } \\
 &= \sum\limits_{R=1}^{N}{\frac{1}{M} Var(\widehat{\mathcal M}(r)|R)} \nonumber
    \end{split}
\end{equation}

\section{Least Square Formulation}\label{sec:app_ls}

This section would take the conclusion from \Cref{sec:app_linear} to derive the quadratic form for \Cref{eq:new_loss}:
\begin{small}
\begin{equation}
    \begin{split}
        \mathcal{L}_1&=\sum\limits_{R=1}^N{P(R)\Big(\sum\limits_{r=1}^n{P(r|R) \widehat{\mathcal M}(r)-\mathcal{M}(R)}\Big)^2} \\
        & = ||\sqrt{D}A\mathbf{x}-\sqrt{D}\mathbf{b}||^2_F\\
        \mathcal{L}_2&=\sum\limits_{R=1}^N{P(R)\cdot \E[\sum_{r=1}^n \tilde{P}(r|R) \widehat{\mathcal M}(r)- \sum_{r=1}^n  P(r|R) \widehat{\mathcal M}(r)]^2 }\\
        &=\sum\limits_{R=1}^N{P(R)\cdot \E[\sum_{r=1}^n \frac{X_r}{M\cdot P(R)} \widehat{\mathcal M}(r)- \sum_{r=1}^n  \frac{\E[X_r]}{M\cdot P(R)} \widehat{\mathcal M}(r)]^2 }\\
        &=\sum\limits_{R=1}^N{\frac{1}{m^2\cdot P(R)}\cdot \E[\sum_{r=1}^n {X_r}\widehat{\mathcal M}(r)- \sum_{r=1}^n  {\E[X_r]} \widehat{\mathcal M}(r)]^2 }\\
 &= \sum\limits_{R=1}^{N}{\frac{1}{M}\cdot \Big(\sum\limits_{r}^n{\widehat{\mathcal M}^2(r) P(r|R)}-\big(\sum\limits_{r}^n{\widehat{\mathcal M}(r)P(r|R)}\big)^2 \Big) }\\
        &=\frac{1}{M}\mathbf{x}^T\Lambda_1 \mathbf{x} - \frac{1}{M}||A\mathbf{x}||^2_F=\frac{1}{M}||\sqrt{\Lambda_1} \mathbf{x}||^2_F - \frac{1}{M}||Ax||^2_F
    \end{split}
\end{equation}
\end{small}

\section{EM steps}\label{sec:app_em}
\noindent{\textbf{E-step}}\\
\begin{equation}
    \begin{split}
        \log\mathcal{L}&=\sum\limits_{u=1}^{M}\log\sum\limits_{k=1}^N{P(r_u, R_{uk}; \boldsymbol{\pi})}\\
        &= \sum\limits_{u=1}^{M}\log\sum\limits_{k=1}^N{\phi(R_{uk})\cdot \frac{P(r_u, R_{uk}; \boldsymbol{\pi})}{\phi(R_{uk})}}\\
       &\ge \sum\limits_{u=1}^{M}\sum\limits_{k=1}^N{\phi(R_{uk})\cdot\log P(r_u, R_{uk}; \boldsymbol{\pi})} + constant\\
       &\triangleq \sum\limits_{u=1}^{M} Q_u(\boldsymbol{\pi}, \boldsymbol{\pi}^{old}) = Q(\boldsymbol{\pi}, \boldsymbol{\pi}^{old})
       \end{split}
       \end{equation}
where
       \begin{equation}
           \begin{split}
        \phi(R_{uk}) &= P(R_u = k|r_u;\boldsymbol{\pi}^{old})\\
        &=\frac{\pi^{old}_k\cdot P(r_u|R_u = k; n_u)}{\sum\limits_{j=1}^N\pi^{old}_j\cdot P(r_u|R_u = j; n_u) }
    \end{split}
\end{equation}
where $\phi$ is the posterior, $\boldsymbol{\pi}^{old}$ is the known probability distribution and $\boldsymbol{\pi}$ is parameter.

\noindent{\textbf{M-step}:} Derived from the Lagrange maximization procedure:  
\begin{equation}
    \begin{split}
        \pi^{new}_k=\frac{1}{M}\sum\limits_{u=1}^M\phi(R_{uk})
    \end{split}
\end{equation}

\section{Adaptive Sampling Explanation}
\label{sec:adp_sampling}
A detailed explanation for \cref{alg:adp}. Specifically, we start from an initial sample set size parameter $n_0$. We sample $n_0-1$ items and compute the rank $r_u$ for all users. For those users with $r_u>1$, we take down the sample set size $n_u=n_0$. For those with $r_u = 1$, we double the sample set size $n_1 = 2n_0$, in other words, we sample another set of $n_0$ items (since we already sample $n_0-1$). Consequently, we check the rank $r_u$ and repeat the process until $r_u\neq 1$ or sample set size is $n_{max}$. We will discuss how to determine $n_{max}$ later in ~\cref{upperbound}.

\section{More Experimental Results}\label{sec:app_exp}

\subsection{Item-sampling based estimators} 
The estimators evaluated in this section are: $\mathbf{BV}$ (Bias-Variance Estimator)\cite{Krichene20@KDD20}; $\mathbf{MLE}$ (Maximal Likelihood Estimation)\cite{Jin@AAAI21}; $\mathbf{MES}$ (Maximal Entropy with Squared distribution distance)\cite{Jin@AAAI21}; 
$\textbf{BV\_MLE}$, $\textbf{BV\_MES}$ (\cref{eq:leastsquare} with $P(R)$ obtained from $MLE$ and $MES$; basically, we consider combining the two approaches from \textbf{BV} \cite{Krichene20@KDD20} and $\mathbf{MLE}$ ($\textbf{MES}$) \cite{Jin@AAAI21}); the new multinomial distribution based estimator with different prior, short as $\textbf{MN\_MLE}$, $\textbf{MN\_MES}$, \cref{eq:mn_closed} with prior $P(R)$ obtained from $MLE$ and $MES$ estimators.

\subsection{Reproducibility}

We release code at \url{https://github.com/dli12/AAAI23-Towards-Reliable-Item-Sampling-for-Recommendation-Evaluation}.

\subsection{Dataset Statistics}

\Cref{tab:datasets} shows the information of the three datasets used in this paper. 
\begin{table}[]
  \caption{Dataset Statistics}
  \label{tab:datasets}
  \begin{tabular}{lcccc}
    \toprule
    \textbf{Dataset}&
    \textbf{Interactions}&
    \textbf{Users}&
    \textbf{Items}&\textbf{Sparsity}\\
    \midrule
    pinterest-20& 1,463,581 & 55,187&9,916&99.73$\%$\\
    yelp& 696,865&25,677 &25,815&99.89$\%$\\
    ml-20m&9,990,682&136,677&20,720&99.65$\%$\\
    \bottomrule
  \end{tabular}
\end{table}

\subsection{Complete Table}
Due to space limitation, we put the complete version of \cref{tab:recall_error_100}, \cref{tab:ndcg_adaptive} and \cref{tab:winner} in below, see \cref{tab:recall_error_100_cp}, \cref{tab:ndcg_adaptive_cp} and \cref{tab:winner_cp} respectively.

\begin{table*}[]
\centering
\caption{The average relative errors between estimated $Recall@K$ ($K$ from $1$ to $50$) and the true ones. Unit is $\%$. In each row, the smallest two results are highlighted in bold, indicating the most accurate results. Sample set size $n=100$.}
\label{tab:recall_error_100_cp}
\resizebox{0.9\textwidth}{!}{%
\begin{tabular}{|c|c|rrrrrrr|}
\hline
\multirow{3}{*}{dataset} &
  \multirow{3}{*}{Models} &
  \multicolumn{7}{c|}{sample set size 100} \\ \cline{3-9} 
 &
   &
  \multicolumn{3}{c|}{baseline} &
  \multicolumn{4}{c|}{this paper} \\ \cline{3-9} 
 &
   &
  \multicolumn{1}{c|}{MES} &
  \multicolumn{1}{c|}{MLE} &
  \multicolumn{1}{c|}{BV} &
  \multicolumn{1}{c|}{BV\_MES} &
  \multicolumn{1}{c|}{BV\_MLE} &
  \multicolumn{1}{c|}{MN\_MES} &
  \multicolumn{1}{c|}{MN\_MLE} \\ \hline
\multirow{5}{*}{pinterest-20} &
  EASE &
  \multicolumn{1}{r|}{5.86$\rpm$2.26} &
  \multicolumn{1}{r|}{5.54$\rpm$1.85} &
  \multicolumn{1}{r|}{8.11$\rpm$2.00} &
  \multicolumn{1}{r|}{\textbf{5.05$\rpm$1.46}} &
  \multicolumn{1}{r|}{5.14$\rpm$1.46} &
  \multicolumn{1}{r|}{\textbf{5.00$\rpm$1.39}} &
  5.10$\rpm$1.34 \\ \cline{2-9} 
 &
  MultiVAE &
  \multicolumn{1}{r|}{4.17$\rpm$2.91} &
  \multicolumn{1}{r|}{3.34$\rpm$2.07} &
  \multicolumn{1}{r|}{\textbf{2.75$\rpm$1.61}} &
  \multicolumn{1}{r|}{2.89$\rpm$1.74} &
  \multicolumn{1}{r|}{\textbf{2.88$\rpm$1.74}} &
  \multicolumn{1}{r|}{\textbf{2.75$\rpm$1.66}} &
  \textbf{2.75$\rpm$1.68} \\ \cline{2-9} 
 &
  NeuMF &
  \multicolumn{1}{r|}{5.17$\rpm$2.74} &
  \multicolumn{1}{r|}{4.28$\rpm$1.95} &
  \multicolumn{1}{r|}{4.23$\rpm$1.79} &
  \multicolumn{1}{r|}{3.83$\rpm$1.59} &
  \multicolumn{1}{r|}{3.84$\rpm$1.72} &
  \multicolumn{1}{r|}{\textbf{3.60$\rpm$1.50}} &
  \textbf{3.76$\rpm$1.44} \\ \cline{2-9} 
 &
  itemKNN &
  \multicolumn{1}{r|}{5.90$\rpm$2.20} &
  \multicolumn{1}{r|}{5.80$\rpm$1.60} &
  \multicolumn{1}{r|}{8.93$\rpm$1.70} &
  \multicolumn{1}{r|}{\textbf{5.11$\rpm$1.22}} &
  \multicolumn{1}{r|}{5.31$\rpm$1.25} &
  \multicolumn{1}{r|}{\textbf{5.09$\rpm$1.15}} &
  5.26$\rpm$1.14 \\ \cline{2-9} 
 &
  ALS &
  \multicolumn{1}{r|}{4.19$\rpm$2.37} &
  \multicolumn{1}{r|}{3.44$\rpm$1.68} &
  \multicolumn{1}{r|}{3.17$\rpm$1.34} &
  \multicolumn{1}{r|}{3.05$\rpm$1.39} &
  \multicolumn{1}{r|}{3.07$\rpm$1.42} &
  \multicolumn{1}{r|}{\textbf{2.86$\rpm$1.27}} &
  \textbf{2.90$\rpm$1.28} \\ \hline
\multirow{5}{*}{yelp} &
  EASE &
  \multicolumn{1}{r|}{8.08$\rpm$4.94} &
  \multicolumn{1}{r|}{7.89$\rpm$4.70} &
  \multicolumn{1}{r|}{18.60$\rpm$2.78} &
  \multicolumn{1}{r|}{6.10$\rpm$3.74} &
  \multicolumn{1}{r|}{6.56$\rpm$3.90} &
  \multicolumn{1}{r|}{\textbf{4.84$\rpm$2.17}} &
  \textbf{5.61$\rpm$2.30} \\ \cline{2-9} 
 &
  MultiVAE &
  \multicolumn{1}{r|}{9.33$\rpm$6.61} &
  \multicolumn{1}{r|}{7.67$\rpm$4.94} &
  \multicolumn{1}{r|}{9.70$\rpm$3.22} &
  \multicolumn{1}{r|}{6.84$\rpm$4.10} &
  \multicolumn{1}{r|}{6.80$\rpm$4.04} &
  \multicolumn{1}{r|}{\textbf{4.30$\rpm$1.27}} &
  \textbf{4.35$\rpm$1.31} \\ \cline{2-9} 
 &
  NeuMF &
  \multicolumn{1}{r|}{15.09$\rpm$6.24} &
  \multicolumn{1}{r|}{15.47$\rpm$5.55} &
  \multicolumn{1}{r|}{22.40$\rpm$3.17} &
  \multicolumn{1}{r|}{\textbf{13.14$\rpm$4.55}} &
  \multicolumn{1}{r|}{13.92$\rpm$4.70} &
  \multicolumn{1}{r|}{\textbf{13.46$\rpm$2.43}} &
  14.50$\rpm$2.45 \\ \cline{2-9} 
 &
  itemKNN &
  \multicolumn{1}{r|}{9.25$\rpm$4.87} &
  \multicolumn{1}{r|}{9.62$\rpm$4.88} &
  \multicolumn{1}{r|}{23.24$\rpm$2.16} &
  \multicolumn{1}{r|}{\textbf{7.69$\rpm$4.09}} &
  \multicolumn{1}{r|}{8.15$\rpm$4.17} &
  \multicolumn{1}{r|}{\textbf{7.74$\rpm$2.08}} &
  8.75$\rpm$2.08 \\ \cline{2-9} 
 &
  ALS &
  \multicolumn{1}{r|}{14.31$\rpm$3.96} &
  \multicolumn{1}{r|}{13.68$\rpm$3.51} &
  \multicolumn{1}{r|}{15.14$\rpm$1.86} &
  \multicolumn{1}{r|}{13.43$\rpm$3.16} &
  \multicolumn{1}{r|}{13.26$\rpm$3.08} &
  \multicolumn{1}{r|}{\textbf{11.68$\rpm$0.88}} &
  \textbf{11.57$\rpm$0.83} \\ \hline
\multirow{5}{*}{ml-20m} &
  EASE &
  \multicolumn{1}{r|}{10.45$\rpm$1.03} &
  \multicolumn{1}{r|}{11.52$\rpm$1.03} &
  \multicolumn{1}{r|}{36.59$\rpm$0.31} &
  \multicolumn{1}{r|}{\textbf{8.99$\rpm$0.74}} &
  \multicolumn{1}{r|}{9.86$\rpm$0.77} &
  \multicolumn{1}{r|}{\textbf{9.07$\rpm$0.61}} &
  10.09$\rpm$0.69 \\ \cline{2-9} 
 &
  MultiVAE &
  \multicolumn{1}{r|}{9.93$\rpm$0.38} &
  \multicolumn{1}{r|}{\textbf{9.48$\rpm$0.22}} &
  \multicolumn{1}{r|}{22.24$\rpm$0.37} &
  \multicolumn{1}{r|}{9.85$\rpm$0.36} &
  \multicolumn{1}{r|}{\textbf{9.50$\rpm$0.22}} &
  \multicolumn{1}{r|}{9.82$\rpm$0.28} &
  9.53$\rpm$0.14 \\ \cline{2-9} 
 &
  NeuMF &
  \multicolumn{1}{r|}{4.35$\rpm$1.50} &
  \multicolumn{1}{r|}{6.05$\rpm$1.35} &
  \multicolumn{1}{r|}{28.27$\rpm$0.42} &
  \multicolumn{1}{r|}{\textbf{3.67$\rpm$1.14}} &
  \multicolumn{1}{r|}{4.81$\rpm$1.14} &
  \multicolumn{1}{r|}{\textbf{3.64$\rpm$1.05}} &
  4.79$\rpm$1.08 \\ \cline{2-9} 
 &
  itemKNN &
  \multicolumn{1}{r|}{15.31$\rpm$1.18} &
  \multicolumn{1}{r|}{17.19$\rpm$1.15} &
  \multicolumn{1}{r|}{36.63$\rpm$0.42} &
  \multicolumn{1}{r|}{\textbf{14.02$\rpm$0.75}} &
  \multicolumn{1}{r|}{15.24$\rpm$0.83} &
  \multicolumn{1}{r|}{\textbf{14.16$\rpm$0.68}} &
  15.41$\rpm$0.77 \\ \cline{2-9} 
 &
  ALS &
  \multicolumn{1}{r|}{36.17$\rpm$0.83} &
  \multicolumn{1}{r|}{\textbf{35.21$\rpm$0.64}} &
  \multicolumn{1}{r|}{36.39$\rpm$0.21} &
  \multicolumn{1}{r|}{36.50$\rpm$0.74} &
  \multicolumn{1}{r|}{35.75$\rpm$0.62} &
  \multicolumn{1}{r|}{36.32$\rpm$0.56} &
  \textbf{35.60$\rpm$0.48} \\ \hline
\end{tabular}%
}
% \end{small}
\end{table*}

\begin{table*}[]
\centering
\caption{The average relative errors between estimated $NDCG@K$ ($K$ from $1$ to $50$) and the true ones. Unit is $\%$. In each row, the smallest relative error is highlighted, indicating the most accurate result.}
\label{tab:ndcg_adaptive_cp}
% \usepackage{multirow}
% \usepackage{graphicx}
\resizebox{0.9\textwidth}{!}{%
\begin{tabular}{|cc|rrrr|rr|}
\hline
\multicolumn{1}{|c|}{\multirow{2}{*}{Dataset}} &
  \multirow{2}{*}{Models} &
  \multicolumn{4}{c|}{Fix Sample} &
  \multicolumn{2}{c|}{Adaptive Sample} \\ \cline{3-8} 
\multicolumn{1}{|c|}{} &
   &
  \multicolumn{1}{c|}{BV\_MES} &
  \multicolumn{1}{c|}{BV\_MLE} &
  \multicolumn{1}{c|}{MN\_MES} &
  \multicolumn{1}{c|}{MN\_MLE} &
  \multicolumn{1}{c|}{average size} &
  \multicolumn{1}{c|}{adaptive MLE} \\ \hline
\multicolumn{2}{|c|}{} &
  \multicolumn{4}{c|}{sample set size n = 500} &
  \multicolumn{2}{c|}{} \\ \hline
\multicolumn{1}{|c|}{\multirow{5}{*}{pinterest-20}} &
  EASE &
  \multicolumn{1}{r|}{3.87$\rpm$2.13} &
  \multicolumn{1}{r|}{4.33$\rpm$2.23} &
  \multicolumn{1}{r|}{4.17$\rpm$2.45} &
  4.33$\rpm$2.50 &
  \multicolumn{1}{r|}{307.74$\rpm$1.41} &
  \textbf{1.46$\rpm$0.63} \\ \cline{2-8} 
\multicolumn{1}{|c|}{} &
  MultiVAE &
  \multicolumn{1}{r|}{2.66$\rpm$1.75} &
  \multicolumn{1}{r|}{2.58$\rpm$1.75} &
  \multicolumn{1}{r|}{3.26$\rpm$2.14} &
  3.07$\rpm$2.09 &
  \multicolumn{1}{r|}{286.46$\rpm$1.48} &
  \textbf{1.67$\rpm$0.70} \\ \cline{2-8} 
\multicolumn{1}{|c|}{} &
  NeuMF &
  \multicolumn{1}{r|}{2.82$\rpm$1.98} &
  \multicolumn{1}{r|}{2.79$\rpm$1.96} &
  \multicolumn{1}{r|}{3.24$\rpm$2.30} &
  3.22$\rpm$2.27 &
  \multicolumn{1}{r|}{259.77$\rpm$1.28} &
  \textbf{1.73$\rpm$0.83} \\ \cline{2-8} 
\multicolumn{1}{|c|}{} &
  itemKNN &
  \multicolumn{1}{r|}{3.99$\rpm$2.22} &
  \multicolumn{1}{r|}{4.18$\rpm$2.24} &
  \multicolumn{1}{r|}{4.23$\rpm$2.47} &
  4.06$\rpm$2.44 &
  \multicolumn{1}{r|}{309.56$\rpm$1.31} &
  \textbf{1.42$\rpm$0.67} \\ \cline{2-8} 
\multicolumn{1}{|c|}{} &
  ALS &
  \multicolumn{1}{r|}{3.35$\rpm$1.89} &
  \multicolumn{1}{r|}{3.24$\rpm$1.87} &
  \multicolumn{1}{r|}{3.90$\rpm$2.24} &
  3.80$\rpm$2.23 &
  \multicolumn{1}{r|}{270.75$\rpm$1.22} &
  \textbf{1.84$\rpm$1.07} \\ \hline
\multicolumn{2}{|c|}{} &
  \multicolumn{4}{c|}{sample set size n = 500} &
  \multicolumn{2}{c|}{} \\ \hline
\multicolumn{1}{|c|}{\multirow{5}{*}{yelp}} &
  EASE &
  \multicolumn{1}{r|}{5.63$\rpm$3.73} &
  \multicolumn{1}{r|}{5.48$\rpm$3.66} &
  \multicolumn{1}{r|}{4.03$\rpm$2.53} &
  4.04$\rpm$2.53 &
  \multicolumn{1}{r|}{340.79$\rpm$2.03} &
  \textbf{3.55$\rpm$2.00} \\ \cline{2-8} 
\multicolumn{1}{|c|}{} &
  MultiVAE &
  \multicolumn{1}{r|}{7.68$\rpm$5.86} &
  \multicolumn{1}{r|}{7.48$\rpm$5.74} &
  \multicolumn{1}{r|}{5.77$\rpm$3.87} &
  5.64$\rpm$3.82 &
  \multicolumn{1}{r|}{288.70$\rpm$2.24} &
  \textbf{5.09$\rpm$2.60} \\ \cline{2-8} 
\multicolumn{1}{|c|}{} &
  NeuMF &
  \multicolumn{1}{r|}{9.34$\rpm$5.50} &
  \multicolumn{1}{r|}{9.55$\rpm$5.50} &
  \multicolumn{1}{r|}{8.43$\rpm$4.07} &
  8.91$\rpm$4.13 &
  \multicolumn{1}{r|}{290.62$\rpm$2.11} &
  \textbf{4.43$\rpm$2.55} \\ \cline{2-8} 
\multicolumn{1}{|c|}{} &
  itemKNN &
  \multicolumn{1}{r|}{5.01$\rpm$2.99} &
  \multicolumn{1}{r|}{4.99$\rpm$2.96} &
  \multicolumn{1}{r|}{\textbf{3.65$\rpm$2.28}} &
  3.71$\rpm$2.29 &
  \multicolumn{1}{r|}{369.16$\rpm$2.51} &
  3.67$\rpm$2.73 \\ \cline{2-8} 
\multicolumn{1}{|c|}{} &
  ALS &
  \multicolumn{1}{r|}{13.39$\rpm$7.34} &
  \multicolumn{1}{r|}{13.94$\rpm$7.61} &
  \multicolumn{1}{r|}{12.57$\rpm$5.46} &
  13.67$\rpm$5.80 &
  \multicolumn{1}{r|}{297.07$\rpm$2.29} &
  \textbf{5.48$\rpm$3.34} \\ \hline
\multicolumn{2}{|c|}{} &
  \multicolumn{4}{c|}{sample set size n = 1000} &
  \multicolumn{2}{c|}{} \\ \hline
\multicolumn{1}{|c|}{\multirow{5}{*}{ml-20m}} &
  EASE &
  \multicolumn{1}{r|}{3.45$\rpm$0.83} &
  \multicolumn{1}{r|}{4.16$\rpm$0.81} &
  \multicolumn{1}{r|}{3.74$\rpm$1.51} &
  3.85$\rpm$1.53 &
  \multicolumn{1}{r|}{899.89$\rpm$1.90} &
  \textbf{2.01$\rpm$0.56} \\ \cline{2-8} 
\multicolumn{1}{|c|}{} &
  MultiVAE &
  \multicolumn{1}{r|}{3.20$\rpm$1.44} &
  \multicolumn{1}{r|}{4.37$\rpm$1.58} &
  \multicolumn{1}{r|}{4.76$\rpm$2.68} &
  3.87$\rpm$2.58 &
  \multicolumn{1}{r|}{771.26$\rpm$1.84} &
  \textbf{1.21$\rpm$0.60} \\ \cline{2-8} 
\multicolumn{1}{|c|}{} &
  NeuMF &
  \multicolumn{1}{r|}{1.02$\rpm$0.65} &
  \multicolumn{1}{r|}{1.14$\rpm$0.74} &
  \multicolumn{1}{r|}{1.94$\rpm$1.35} &
  1.90$\rpm$1.23 &
  \multicolumn{1}{r|}{758.45$\rpm$1.61} &
  \textbf{0.92$\rpm$0.51} \\ \cline{2-8} 
\multicolumn{1}{|c|}{} &
  itemKNN &
  \multicolumn{1}{r|}{4.44$\rpm$1.07} &
  \multicolumn{1}{r|}{5.36$\rpm$1.05} &
  \multicolumn{1}{r|}{3.52$\rpm$1.67} &
  4.13$\rpm$1.68 &
  \multicolumn{1}{r|}{725.72$\rpm$1.49} &
  \textbf{2.02$\rpm$0.63} \\ \cline{2-8} 
\multicolumn{1}{|c|}{} &
  ALS &
  \multicolumn{1}{r|}{12.28$\rpm$2.07} &
  \multicolumn{1}{r|}{17.12$\rpm$2.38} &
  \multicolumn{1}{r|}{13.42$\rpm$4.44} &
  13.48$\rpm$4.90 &
  \multicolumn{1}{r|}{705.76$\rpm$1.56} &
  \textbf{5.27$\rpm$1.39} \\ \hline
\end{tabular}%
}
\end{table*}

\begin{table*}[]
\centering
\caption{Accuracy of predicting of the winner models for different datasets. Values in table is the number of correct predictions over 100 repeats. The larger number, the better estimator.}
\label{tab:winner_cp}
\resizebox{0.8\textwidth}{!}{%
\begin{tabular}{|ccc|cccc|c|}
\hline
\multicolumn{1}{|c|}{\multirow{3}{*}{Dataset}} &
  \multicolumn{1}{c|}{\multirow{3}{*}{Top-K}} &
  \multirow{3}{*}{Metrics} &
  \multicolumn{4}{c|}{Fix Sample} &
  AdaptiveSample \\ \cline{4-8} 
\multicolumn{1}{|c|}{} &
  \multicolumn{1}{c|}{} &
   &
  \multicolumn{1}{c|}{BV\_MES} &
  \multicolumn{1}{c|}{BV\_MLE} &
  \multicolumn{1}{c|}{MN\_MES} &
  MN\_MLE &
  adaptive MLE \\ \cline{4-8} 
\multicolumn{1}{|c|}{} &
  \multicolumn{1}{c|}{} &
   &
  \multicolumn{4}{c|}{sample set size n = 500} &
  sample set size $260\sim 310$ \\ \hline
\multicolumn{1}{|c|}{\multirow{6}{*}{pinterest-20}} &
  \multicolumn{1}{c|}{\multirow{3}{*}{10}} &
  RECALL &
  \multicolumn{1}{c|}{69} &
  \multicolumn{1}{c|}{73} &
  \multicolumn{1}{c|}{67} &
  69 &
  78 \\ \cline{3-8} 
\multicolumn{1}{|c|}{} &
  \multicolumn{1}{c|}{} &
  NDCG &
  \multicolumn{1}{c|}{58} &
  \multicolumn{1}{c|}{59} &
  \multicolumn{1}{c|}{58} &
  60 &
  84 \\ \cline{3-8} 
\multicolumn{1}{|c|}{} &
  \multicolumn{1}{c|}{} &
  AP &
  \multicolumn{1}{c|}{54} &
  \multicolumn{1}{c|}{57} &
  \multicolumn{1}{c|}{52} &
  52 &
  68 \\ \cline{2-8} 
\multicolumn{1}{|c|}{} &
  \multicolumn{1}{c|}{\multirow{3}{*}{20}} &
  RECALL &
  \multicolumn{1}{c|}{69} &
  \multicolumn{1}{c|}{73} &
  \multicolumn{1}{c|}{70} &
  74 &
  81 \\ \cline{3-8} 
\multicolumn{1}{|c|}{} &
  \multicolumn{1}{c|}{} &
  NDCG &
  \multicolumn{1}{c|}{69} &
  \multicolumn{1}{c|}{73} &
  \multicolumn{1}{c|}{68} &
  73 &
  79 \\ \cline{3-8} 
\multicolumn{1}{|c|}{} &
  \multicolumn{1}{c|}{} &
  AP &
  \multicolumn{1}{c|}{57} &
  \multicolumn{1}{c|}{60} &
  \multicolumn{1}{c|}{54} &
  56 &
  69 \\ \hline
\multicolumn{3}{|c|}{} &
  \multicolumn{4}{c|}{sample set size n = 500} &
  sample set size $280\sim 370$ \\ \hline
\multicolumn{1}{|c|}{\multirow{6}{*}{yelp}} &
  \multicolumn{1}{c|}{\multirow{3}{*}{10}} &
  RECALL &
  \multicolumn{1}{c|}{98} &
  \multicolumn{1}{c|}{97} &
  \multicolumn{1}{c|}{100} &
  100 &
  100 \\ \cline{3-8} 
\multicolumn{1}{|c|}{} &
  \multicolumn{1}{c|}{} &
  NDCG &
  \multicolumn{1}{c|}{96} &
  \multicolumn{1}{c|}{96} &
  \multicolumn{1}{c|}{100} &
  100 &
  100 \\ \cline{3-8} 
\multicolumn{1}{|c|}{} &
  \multicolumn{1}{c|}{} &
  AP &
  \multicolumn{1}{c|}{95} &
  \multicolumn{1}{c|}{95} &
  \multicolumn{1}{c|}{100} &
  100 &
  94 \\ \cline{2-8} 
\multicolumn{1}{|c|}{} &
  \multicolumn{1}{c|}{\multirow{3}{*}{20}} &
  RECALL &
  \multicolumn{1}{c|}{100} &
  \multicolumn{1}{c|}{100} &
  \multicolumn{1}{c|}{100} &
  100 &
  100 \\ \cline{3-8} 
\multicolumn{1}{|c|}{} &
  \multicolumn{1}{c|}{} &
  NDCG &
  \multicolumn{1}{c|}{100} &
  \multicolumn{1}{c|}{100} &
  \multicolumn{1}{c|}{100} &
  100 &
  100 \\ \cline{3-8} 
\multicolumn{1}{|c|}{} &
  \multicolumn{1}{c|}{} &
  AP &
  \multicolumn{1}{c|}{97} &
  \multicolumn{1}{c|}{96} &
  \multicolumn{1}{c|}{100} &
  100 &
  98 \\ \hline
\multicolumn{3}{|c|}{} &
  \multicolumn{4}{c|}{sample set size n = 1000} &
  sample set size $700\sim 900$ \\ \hline
\multicolumn{1}{|c|}{\multirow{6}{*}{ml-20m}} &
  \multicolumn{1}{c|}{\multirow{3}{*}{10}} &
  RECALL &
  \multicolumn{1}{c|}{100} &
  \multicolumn{1}{c|}{100} &
  \multicolumn{1}{c|}{100} &
  100 &
  100 \\ \cline{3-8} 
\multicolumn{1}{|c|}{} &
  \multicolumn{1}{c|}{} &
  NDCG &
  \multicolumn{1}{c|}{100} &
  \multicolumn{1}{c|}{100} &
  \multicolumn{1}{c|}{100} &
  100 &
  100 \\ \cline{3-8} 
\multicolumn{1}{|c|}{} &
  \multicolumn{1}{c|}{} &
  AP &
  \multicolumn{1}{c|}{100} &
  \multicolumn{1}{c|}{100} &
  \multicolumn{1}{c|}{100} &
  100 &
  100 \\ \cline{2-8} 
\multicolumn{1}{|c|}{} &
  \multicolumn{1}{c|}{\multirow{3}{*}{20}} &
  RECALL &
  \multicolumn{1}{c|}{100} &
  \multicolumn{1}{c|}{100} &
  \multicolumn{1}{c|}{100} &
  100 &
  100 \\ \cline{3-8} 
\multicolumn{1}{|c|}{} &
  \multicolumn{1}{c|}{} &
  NDCG &
  \multicolumn{1}{c|}{100} &
  \multicolumn{1}{c|}{100} &
  \multicolumn{1}{c|}{100} &
  100 &
  100 \\ \cline{3-8} 
\multicolumn{1}{|c|}{} &
  \multicolumn{1}{c|}{} &
  AP &
  \multicolumn{1}{c|}{100} &
  \multicolumn{1}{c|}{100} &
  \multicolumn{1}{c|}{100} &
  100 &
  100 \\ \hline
\end{tabular}%
}
\end{table*}

\subsection{More Results}

% \usepackage{multirow}
% \usepackage{graphicx}
Due to the space limitation, we present two complementary experimental results for \cref{tab:recall_error_100} as well as \cref{tab:ndcg_adaptive}.

\Cref{tab:ap_error_100} show the result of average relative errors between estimated $AP@K$ ($K$ from $1$ to $50$) and the true ones. The proposed estimator especially $MN\_MES$ performs best in general, which is consistent with the observation in \Cref{sec:exp}.

\Cref{tab:ndcg_error_500} show the result of average relative errors between estimated $NDCG@K$ ($K$ from $1$ to $50$) and the true ones. In \Cref{tab:recall_error_100}, the sample set size $n=100$ while here is $500$.
The result consistently confirm the superiority of models proposed in this paper. Specifically, from the \Cref{tab:ndcg_error_500}, we can find that $MN\_MES$ estimator performs best or second best (highlighted) in most of the cases.

\Cref{tab:recall_adaptive} presents the average relative errors in terms of $Recall$ metric. There are four advanced estimators with fixed sampling set size ($n=500$ or $1000$) and a proposed adaptive based estimator. We follow the adaptive sampling strategy in \Cref{sec:adaptive} and obtain the average sampling set size for each dataset and model. For example, the average sampling set size is $307$ for EASE model in pinterest-20 dataset, compared to the estimators with $n=500$ sample set size. Nevertheless, the adaptive based estimator can even achive much better performance, say $1.69\%$ relative error compared to the $2.78\%$ from MN\_MES estimator. All these results consistently support the superiority of the adaptive sampling strategy and the adaptive estimator.

\begin{table*}[]
\centering
\caption{The average relative errors between estimated $AP@K$ ($K$ from $1$ to $50$) and the true ones. Unit is $\%$. In each row, the smallest two results are highlighted in bold, indicating the most accurate results. Sample set size $n=100$.}
\label{tab:ap_error_100}
% \usepackage{multirow}
% \usepackage{graphicx}
\resizebox{0.9\textwidth}{!}{%
\begin{tabular}{|c|c|lllllll|}
\hline
\multirow{3}{*}{dataset} &
  \multirow{3}{*}{Models} &
  \multicolumn{7}{c|}{sample set size 100} \\ \cline{3-9} 
 &
   &
  \multicolumn{3}{c|}{baseline} &
  \multicolumn{4}{c|}{this paper} \\ \cline{3-9} 
 &
   &
  \multicolumn{1}{c|}{MES} &
  \multicolumn{1}{c|}{MLE} &
  \multicolumn{1}{c|}{BV} &
  \multicolumn{1}{c|}{BV\_MES} &
  \multicolumn{1}{c|}{BV\_MLE} &
  \multicolumn{1}{c|}{MN\_MES} &
  \multicolumn{1}{c|}{MN\_MLE} \\ \hline
\multirow{5}{*}{pinterest-20} &
  EASE &
  \multicolumn{1}{l|}{17.45$\rpm$6.14} &
  \multicolumn{1}{l|}{17.85$\rpm$4.50} &
  \multicolumn{1}{l|}{23.50$\rpm$2.25} &
  \multicolumn{1}{l|}{\textbf{16.88$\rpm$3.91}} &
  \multicolumn{1}{l|}{17.19$\rpm$3.87} &
  \multicolumn{1}{l|}{\textbf{17.06$\rpm$3.54}} &
  17.30$\rpm$3.52 \\ \cline{2-9} 
 &
  MultiVAE &
  \multicolumn{1}{l|}{6.45$\rpm$5.05} &
  \multicolumn{1}{l|}{4.68$\rpm$3.40} &
  \multicolumn{1}{l|}{\textbf{3.58$\rpm$2.29}} &
  \multicolumn{1}{l|}{4.04$\rpm$2.94} &
  \multicolumn{1}{l|}{4.03$\rpm$2.93} &
  \multicolumn{1}{l|}{\textbf{3.78$\rpm$2.66}} &
  \textbf{3.76$\rpm$2.66} \\ \cline{2-9} 
 &
  NeuMF &
  \multicolumn{1}{l|}{8.62$\rpm$6.12} &
  \multicolumn{1}{l|}{7.28$\rpm$4.77} &
  \multicolumn{1}{l|}{9.43$\rpm$3.33} &
  \multicolumn{1}{l|}{\textbf{6.39$\rpm$4.19}} &
  \multicolumn{1}{l|}{7.04$\rpm$4.39} &
  \multicolumn{1}{l|}{\textbf{6.19$\rpm$3.89}} &
  \textbf{6.58$\rpm$3.99} \\ \cline{2-9} 
 &
  itemKNN &
  \multicolumn{1}{l|}{17.46$\rpm$6.07} &
  \multicolumn{1}{l|}{18.17$\rpm$4.12} &
  \multicolumn{1}{l|}{23.90$\rpm$1.97} &
  \multicolumn{1}{l|}{\textbf{16.81$\rpm$3.53}} &
  \multicolumn{1}{l|}{17.41$\rpm$3.49} &
  \multicolumn{1}{l|}{\textbf{17.02$\rpm$3.17}} &
  17.46$\rpm$3.15 \\ \cline{2-9} 
 &
  ALS &
  \multicolumn{1}{l|}{7.78$\rpm$5.56} &
  \multicolumn{1}{l|}{7.03$\rpm$4.30} &
  \multicolumn{1}{l|}{\textbf{3.39$\rpm$1.79}} &
  \multicolumn{1}{l|}{6.31$\rpm$3.70} &
  \multicolumn{1}{l|}{6.48$\rpm$3.76} &
  \multicolumn{1}{l|}{\textbf{5.80$\rpm$3.30}} &
  6.07$\rpm$3.39 \\ \hline
\multirow{5}{*}{yelp} &
  EASE &
  \multicolumn{1}{l|}{9.87$\rpm$6.09} &
  \multicolumn{1}{l|}{9.72$\rpm$5.78} &
  \multicolumn{1}{l|}{22.16$\rpm$2.90} &
  \multicolumn{1}{l|}{7.44$\rpm$4.68} &
  \multicolumn{1}{l|}{8.04$\rpm$4.87} &
  \multicolumn{1}{l|}{\textbf{6.14$\rpm$2.64}} &
  \textbf{7.22$\rpm$2.72} \\ \cline{2-9} 
 &
  MultiVAE &
  \multicolumn{1}{l|}{12.07$\rpm$10.12} &
  \multicolumn{1}{l|}{9.66$\rpm$7.90} &
  \multicolumn{1}{l|}{\textbf{6.00$\rpm$3.04}} &
  \multicolumn{1}{l|}{9.53$\rpm$7.03} &
  \multicolumn{1}{l|}{9.11$\rpm$6.82} &
  \multicolumn{1}{l|}{\textbf{6.85$\rpm$3.24}} &
  \textbf{6.20$\rpm$3.08} \\ \cline{2-9} 
 &
  NeuMF &
  \multicolumn{1}{l|}{33.32$\rpm$7.85} &
  \multicolumn{1}{l|}{34.41$\rpm$6.23} &
  \multicolumn{1}{l|}{41.36$\rpm$2.64} &
  \multicolumn{1}{l|}{\textbf{32.00$\rpm$5.34}} &
  \multicolumn{1}{l|}{\textbf{33.04$\rpm$5.23}} &
  \multicolumn{1}{l|}{33.05$\rpm$2.14} &
  34.15$\rpm$2.12 \\ \cline{2-9} 
 &
  itemKNN &
  \multicolumn{1}{l|}{14.97$\rpm$6.82} &
  \multicolumn{1}{l|}{15.94$\rpm$6.24} &
  \multicolumn{1}{l|}{31.18$\rpm$2.11} &
  \multicolumn{1}{l|}{\textbf{13.43$\rpm$5.48}} &
  \multicolumn{1}{l|}{\textbf{14.17$\rpm$5.43}} &
  \multicolumn{1}{l|}{\textbf{14.21$\rpm$2.20}} &
  15.45$\rpm$2.17 \\ \cline{2-9} 
 &
  ALS &
  \multicolumn{1}{l|}{31.31$\rpm$13.11} &
  \multicolumn{1}{l|}{\textbf{31.06$\rpm$11.63}} &
  \multicolumn{1}{l|}{\textbf{13.59$\rpm$4.14}} &
  \multicolumn{1}{l|}{32.91$\rpm$10.21} &
  \multicolumn{1}{l|}{32.38$\rpm$10.10} &
  \multicolumn{1}{l|}{32.00$\rpm$4.18} &
  \textbf{31.29$\rpm$4.17} \\ \hline
\multirow{5}{*}{ml-20m} &
  EASE &
  \multicolumn{1}{l|}{31.30$\rpm$1.29} &
  \multicolumn{1}{l|}{32.83$\rpm$1.23} &
  \multicolumn{1}{l|}{57.14$\rpm$0.23} &
  \multicolumn{1}{l|}{\textbf{29.29$\rpm$1.10}} &
  \multicolumn{1}{l|}{30.66$\rpm$1.05} &
  \multicolumn{1}{l|}{\textbf{29.48$\rpm$0.91}} &
  31.04$\rpm$0.90 \\ \cline{2-9} 
 &
  MultiVAE &
  \multicolumn{1}{l|}{29.01$\rpm$2.90} &
  \multicolumn{1}{l|}{\textbf{25.61$\rpm$2.54}} &
  \multicolumn{1}{l|}{\textbf{12.75$\rpm$0.42}} &
  \multicolumn{1}{l|}{30.07$\rpm$2.25} &
  \multicolumn{1}{l|}{27.17$\rpm$2.18} &
  \multicolumn{1}{l|}{30.33$\rpm$2.00} &
  \textbf{26.74$\rpm$1.97} \\ \cline{2-9} 
 &
  NeuMF &
  \multicolumn{1}{l|}{9.10$\rpm$2.05} &
  \multicolumn{1}{l|}{11.64$\rpm$1.74} &
  \multicolumn{1}{l|}{37.40$\rpm$0.40} &
  \multicolumn{1}{l|}{\textbf{8.12$\rpm$1.55}} &
  \multicolumn{1}{l|}{9.86$\rpm$1.50} &
  \multicolumn{1}{l|}{\textbf{8.29$\rpm$1.40}} &
  9.99$\rpm$1.37 \\ \cline{2-9} 
 &
  itemKNN &
  \multicolumn{1}{l|}{43.81$\rpm$1.44} &
  \multicolumn{1}{l|}{46.09$\rpm$1.17} &
  \multicolumn{1}{l|}{62.55$\rpm$0.28} &
  \multicolumn{1}{l|}{\textbf{42.14$\rpm$1.04}} &
  \multicolumn{1}{l|}{43.82$\rpm$1.00} &
  \multicolumn{1}{l|}{\textbf{42.36$\rpm$0.94}} &
  44.12$\rpm$0.92 \\ \cline{2-9} 
 &
  ALS &
  \multicolumn{1}{l|}{88.83$\rpm$4.86} &
  \multicolumn{1}{l|}{\textbf{84.37$\rpm$4.40}} &
  \multicolumn{1}{l|}{\textbf{28.12$\rpm$0.97}} &
  \multicolumn{1}{l|}{91.43$\rpm$3.89} &
  \multicolumn{1}{l|}{87.73$\rpm$3.77} &
  \multicolumn{1}{l|}{90.49$\rpm$3.43} &
  87.04$\rpm$3.38 \\ \hline
\end{tabular}%
}
\end{table*}

\begin{table*}[]
\centering
\caption{The average relative errors between estimated $NDCG@K$ ($K$ from $1$ to $50$) and the true ones. Unit is $\%$. In each row, the smallest two results are highlighted in bold, indicating the most accurate results. Sample set size $n=500$.}
\label{tab:ndcg_error_500}
% \usepackage{multirow}
% \usepackage{graphicx}
\resizebox{0.9\textwidth}{!}{%
\begin{tabular}{|c|c|lllllll|}
\hline
\multirow{3}{*}{dataset} &
  \multirow{3}{*}{Models} &
  \multicolumn{7}{c|}{sample set size 500} \\ \cline{3-9} 
 &
   &
  \multicolumn{3}{c|}{baseline} &
  \multicolumn{4}{c|}{this paper} \\ \cline{3-9} 
 &
   &
  \multicolumn{1}{c|}{MES} &
  \multicolumn{1}{c|}{MLE} &
  \multicolumn{1}{c|}{BV} &
  \multicolumn{1}{c|}{BV\_MES} &
  \multicolumn{1}{c|}{BV\_MLE} &
  \multicolumn{1}{c|}{MN\_MES} &
  \multicolumn{1}{c|}{MN\_MLE} \\ \hline
\multirow{5}{*}{pinterest-20} &
  EASE &
  \multicolumn{1}{l|}{5.11$\rpm$3.09} &
  \multicolumn{1}{l|}{4.52$\rpm$2.42} &
  \multicolumn{1}{l|}{5.53$\rpm$2.12} &
  \multicolumn{1}{l|}{\textbf{3.87$\rpm$2.13}} &
  \multicolumn{1}{l|}{4.33$\rpm$2.23} &
  \multicolumn{1}{l|}{\textbf{4.17$\rpm$2.45}} &
  4.33$\rpm$2.50 \\ \cline{2-9} 
 &
  MultiVAE &
  \multicolumn{1}{l|}{4.01$\rpm$2.60} &
  \multicolumn{1}{l|}{2.89$\rpm$2.05} &
  \multicolumn{1}{l|}{\textbf{2.29$\rpm$1.58}} &
  \multicolumn{1}{l|}{2.66$\rpm$1.75} &
  \multicolumn{1}{l|}{\textbf{2.58$\rpm$1.75}} &
  \multicolumn{1}{l|}{3.26$\rpm$2.14} &
  3.07$\rpm$2.09 \\ \cline{2-9} 
 &
  NeuMF &
  \multicolumn{1}{l|}{4.27$\rpm$2.85} &
  \multicolumn{1}{l|}{3.20$\rpm$2.22} &
  \multicolumn{1}{l|}{\textbf{2.75$\rpm$1.83}} &
  \multicolumn{1}{l|}{2.82$\rpm$1.98} &
  \multicolumn{1}{l|}{\textbf{2.79$\rpm$1.96}} &
  \multicolumn{1}{l|}{3.24$\rpm$2.30} &
  3.22$\rpm$2.27 \\ \cline{2-9} 
 &
  itemKNN &
  \multicolumn{1}{l|}{5.13$\rpm$2.99} &
  \multicolumn{1}{l|}{4.36$\rpm$2.46} &
  \multicolumn{1}{l|}{5.23$\rpm$2.15} &
  \multicolumn{1}{l|}{\textbf{3.99$\rpm$2.22}} &
  \multicolumn{1}{l|}{4.18$\rpm$2.24} &
  \multicolumn{1}{l|}{4.23$\rpm$2.47} &
  \textbf{4.06$\rpm$2.44} \\ \cline{2-9} 
 &
  ALS &
  \multicolumn{1}{l|}{4.53$\rpm$2.86} &
  \multicolumn{1}{l|}{3.65$\rpm$2.13} &
  \multicolumn{1}{l|}{\textbf{3.16$\rpm$1.77}} &
  \multicolumn{1}{l|}{3.35$\rpm$1.89} &
  \multicolumn{1}{l|}{\textbf{3.24$\rpm$1.87}} &
  \multicolumn{1}{l|}{3.90$\rpm$2.24} &
  3.80$\rpm$2.23 \\ \hline
\multirow{5}{*}{yelp} &
  EASE &
  \multicolumn{1}{l|}{8.59$\rpm$5.62} &
  \multicolumn{1}{l|}{6.32$\rpm$4.28} &
  \multicolumn{1}{l|}{4.85$\rpm$3.16} &
  \multicolumn{1}{l|}{5.63$\rpm$3.73} &
  \multicolumn{1}{l|}{5.48$\rpm$3.66} &
  \multicolumn{1}{l|}{\textbf{4.03$\rpm$2.53}} &
  \textbf{4.04$\rpm$2.53} \\ \cline{2-9} 
 &
  MultiVAE &
  \multicolumn{1}{l|}{10.83$\rpm$8.83} &
  \multicolumn{1}{l|}{8.57$\rpm$6.91} &
  \multicolumn{1}{l|}{6.48$\rpm$4.79} &
  \multicolumn{1}{l|}{7.68$\rpm$5.86} &
  \multicolumn{1}{l|}{7.48$\rpm$5.74} &
  \multicolumn{1}{l|}{\textbf{5.77$\rpm$3.87}} &
  \textbf{5.64$\rpm$3.82} \\ \cline{2-9} 
 &
  NeuMF &
  \multicolumn{1}{l|}{11.52$\rpm$7.27} &
  \multicolumn{1}{l|}{10.82$\rpm$6.16} &
  \multicolumn{1}{l|}{12.82$\rpm$5.17} &
  \multicolumn{1}{l|}{9.34$\rpm$5.50} &
  \multicolumn{1}{l|}{9.55$\rpm$5.50} &
  \multicolumn{1}{l|}{\textbf{8.43$\rpm$4.07}} &
  \textbf{8.91$\rpm$4.13} \\ \cline{2-9} 
 &
  itemKNN &
  \multicolumn{1}{l|}{7.72$\rpm$5.18} &
  \multicolumn{1}{l|}{5.87$\rpm$3.38} &
  \multicolumn{1}{l|}{5.31$\rpm$3.26} &
  \multicolumn{1}{l|}{5.01$\rpm$2.99} &
  \multicolumn{1}{l|}{4.99$\rpm$2.96} &
  \multicolumn{1}{l|}{\textbf{3.65$\rpm$2.28}} &
  \textbf{3.71$\rpm$2.29} \\ \cline{2-9} 
 &
  ALS &
  \multicolumn{1}{l|}{16.30$\rpm$10.72} &
  \multicolumn{1}{l|}{15.65$\rpm$9.30} &
  \multicolumn{1}{l|}{15.26$\rpm$7.00} &
  \multicolumn{1}{l|}{\textbf{13.39$\rpm$7.34}} &
  \multicolumn{1}{l|}{13.94$\rpm$7.61} &
  \multicolumn{1}{l|}{\textbf{12.57$\rpm$5.46}} &
  13.67$\rpm$5.80 \\ \hline
\multirow{5}{*}{ml-20m} &
  EASE &
  \multicolumn{1}{l|}{5.93$\rpm$2.02} &
  \multicolumn{1}{l|}{7.44$\rpm$1.08} &
  \multicolumn{1}{l|}{14.35$\rpm$0.56} &
  \multicolumn{1}{l|}{5.78$\rpm$0.95} &
  \multicolumn{1}{l|}{6.78$\rpm$0.93} &
  \multicolumn{1}{l|}{\textbf{5.59$\rpm$1.49}} &
  \textbf{6.14$\rpm$1.47} \\ \cline{2-9} 
 &
  MultiVAE &
  \multicolumn{1}{l|}{\textbf{7.53$\rpm$2.95}} &
  \multicolumn{1}{l|}{10.27$\rpm$1.68} &
  \multicolumn{1}{l|}{10.41$\rpm$1.08} &
  \multicolumn{1}{l|}{7.22$\rpm$1.42} &
  \multicolumn{1}{l|}{9.35$\rpm$1.46} &
  \multicolumn{1}{l|}{\textbf{7.01$\rpm$2.18}} &
  8.08$\rpm$2.26 \\ \cline{2-9} 
 &
  NeuMF &
  \multicolumn{1}{l|}{2.82$\rpm$1.92} &
  \multicolumn{1}{l|}{\textbf{1.71$\rpm$1.00}} &
  \multicolumn{1}{l|}{4.41$\rpm$0.91} &
  \multicolumn{1}{l|}{\textbf{1.29$\rpm$0.71}} &
  \multicolumn{1}{l|}{1.46$\rpm$0.83} &
  \multicolumn{1}{l|}{2.10$\rpm$1.31} &
  2.11$\rpm$1.32 \\ \cline{2-9} 
 &
  itemKNN &
  \multicolumn{1}{l|}{8.45$\rpm$2.12} &
  \multicolumn{1}{l|}{10.50$\rpm$1.27} &
  \multicolumn{1}{l|}{18.98$\rpm$0.66} &
  \multicolumn{1}{l|}{\textbf{8.24$\rpm$1.12}} &
  \multicolumn{1}{l|}{9.46$\rpm$1.10} &
  \multicolumn{1}{l|}{\textbf{7.40$\rpm$1.60}} &
  8.96$\rpm$1.56 \\ \cline{2-9} 
 &
  ALS &
  \multicolumn{1}{l|}{28.45$\rpm$5.33} &
  \multicolumn{1}{l|}{37.22$\rpm$2.88} &
  \multicolumn{1}{l|}{40.95$\rpm$1.72} &
  \multicolumn{1}{l|}{\textbf{27.89$\rpm$2.30}} &
  \multicolumn{1}{l|}{33.53$\rpm$2.47} &
  \multicolumn{1}{l|}{\textbf{26.31$\rpm$3.81}} &
  30.61$\rpm$4.08 \\ \hline
\end{tabular}%
}
\end{table*}

\begin{table*}[]
\centering
\caption{The average relative errors between estimated $Recall@K$ ($K$ from $1$ to $50$) and the true ones. Unit is $\%$. In each row, the smallest relative error is highlighted, indicating the most accurate result.}
\label{tab:recall_adaptive}
% \usepackage{multirow}
% \usepackage{graphicx}
\resizebox{0.9\textwidth}{!}{%
\begin{tabular}{|cc|llll|cc|}
\hline
\multicolumn{1}{|c|}{\multirow{2}{*}{Dataset}} &
  \multirow{2}{*}{Models} &
  \multicolumn{4}{c|}{Fix Sample} &
  \multicolumn{2}{c|}{Adaptive Sample} \\ \cline{3-8} 
\multicolumn{1}{|c|}{} &
   &
  \multicolumn{1}{c|}{BV\_MES} &
  \multicolumn{1}{c|}{BV\_MLE} &
  \multicolumn{1}{c|}{MN\_MES} &
  \multicolumn{1}{c|}{MN\_MLE} &
  \multicolumn{1}{c|}{average size} &
  adaptive MLE \\ \hline
\multicolumn{2}{|c|}{} &
  \multicolumn{4}{c|}{sample set size n = 500} &
  \multicolumn{2}{c|}{} \\ \hline
\multicolumn{1}{|c|}{\multirow{5}{*}{pinterest-20}} &
  EASE &
  \multicolumn{1}{l|}{2.54$\rpm$0.85} &
  \multicolumn{1}{l|}{2.68$\rpm$0.87} &
  \multicolumn{1}{l|}{2.78$\rpm$1.05} &
  2.83$\rpm$1.06 &
  \multicolumn{1}{c|}{307.74$\rpm$1.41} &
  \textbf{1.69$\rpm$0.60} \\ \cline{2-8} 
\multicolumn{1}{|c|}{} &
  MultiVAE &
  \multicolumn{1}{l|}{2.17$\rpm$1.08} &
  \multicolumn{1}{l|}{2.13$\rpm$1.09} &
  \multicolumn{1}{l|}{2.60$\rpm$1.30} &
  2.55$\rpm$1.35 &
  \multicolumn{1}{c|}{286.46$\rpm$1.48} &
  \textbf{1.95$\rpm$0.65} \\ \cline{2-8} 
\multicolumn{1}{|c|}{} &
  NeuMF &
  \multicolumn{1}{l|}{2.45$\rpm$1.15} &
  \multicolumn{1}{l|}{2.44$\rpm$1.15} &
  \multicolumn{1}{l|}{2.76$\rpm$1.37} &
  2.80$\rpm$1.38 &
  \multicolumn{1}{c|}{259.77$\rpm$1.28} &
  \textbf{2.00$\rpm$0.81} \\ \cline{2-8} 
\multicolumn{1}{|c|}{} &
  itemKNN &
  \multicolumn{1}{l|}{2.49$\rpm$0.97} &
  \multicolumn{1}{l|}{2.59$\rpm$0.94} &
  \multicolumn{1}{l|}{2.79$\rpm$1.12} &
  2.79$\rpm$1.20 &
  \multicolumn{1}{c|}{309.56$\rpm$1.31} &
  \textbf{1.63$\rpm$0.51} \\ \cline{2-8} 
\multicolumn{1}{|c|}{} &
  ALS &
  \multicolumn{1}{l|}{2.65$\rpm$1.04} &
  \multicolumn{1}{l|}{2.63$\rpm$1.06} &
  \multicolumn{1}{l|}{3.02$\rpm$1.32} &
  2.98$\rpm$1.33 &
  \multicolumn{1}{c|}{270.75$\rpm$1.22} &
  \textbf{2.00$\rpm$0.73} \\ \hline
\multicolumn{2}{|c|}{} &
  \multicolumn{4}{c|}{sample set size n = 500} &
  \multicolumn{2}{c|}{} \\ \hline
\multicolumn{1}{|c|}{\multirow{5}{*}{yelp}} &
  EASE &
  \multicolumn{1}{l|}{4.68$\rpm$2.43} &
  \multicolumn{1}{l|}{4.56$\rpm$2.35} &
  \multicolumn{1}{l|}{\textbf{3.47$\rpm$1.79}} &
  3.49$\rpm$1.78 &
  \multicolumn{1}{c|}{340.79$\rpm$2.03} &
  3.48$\rpm$1.40 \\ \cline{2-8} 
\multicolumn{1}{|c|}{} &
  MultiVAE &
  \multicolumn{1}{l|}{6.14$\rpm$3.48} &
  \multicolumn{1}{l|}{6.07$\rpm$3.46} &
  \multicolumn{1}{l|}{4.68$\rpm$2.27} &
  \textbf{4.67$\rpm$2.28} &
  \multicolumn{1}{c|}{288.70$\rpm$2.24} &
  5.08$\rpm$2.14 \\ \cline{2-8} 
\multicolumn{1}{|c|}{} &
  NeuMF &
  \multicolumn{1}{l|}{6.59$\rpm$2.38} &
  \multicolumn{1}{l|}{6.73$\rpm$2.35} &
  \multicolumn{1}{l|}{5.48$\rpm$1.43} &
  5.68$\rpm$1.42 &
  \multicolumn{1}{c|}{290.62$\rpm$2.11} &
  \textbf{4.01$\rpm$1.51} \\ \cline{2-8} 
\multicolumn{1}{|c|}{} &
  itemKNN &
  \multicolumn{1}{l|}{3.94$\rpm$1.94} &
  \multicolumn{1}{l|}{3.95$\rpm$1.92} &
  \multicolumn{1}{l|}{\textbf{2.92$\rpm$1.60}} &
  2.96$\rpm$1.57 &
  \multicolumn{1}{c|}{369.16$\rpm$2.51} &
  3.25$\rpm$1.59 \\ \cline{2-8} 
\multicolumn{1}{|c|}{} &
  ALS &
  \multicolumn{1}{l|}{10.00$\rpm$3.47} &
  \multicolumn{1}{l|}{10.31$\rpm$3.65} &
  \multicolumn{1}{l|}{9.29$\rpm$2.03} &
  9.80$\rpm$2.23 &
  \multicolumn{1}{c|}{297.07$\rpm$2.29} &
  \textbf{5.25$\rpm$2.38} \\ \hline
\multicolumn{2}{|c|}{} &
  \multicolumn{4}{c|}{sample set size n = 1000} &
  \multicolumn{2}{c|}{} \\ \hline
\multicolumn{1}{|c|}{\multirow{5}{*}{ml-20m}} &
  EASE &
  \multicolumn{1}{l|}{1.39$\rpm$0.21} &
  \multicolumn{1}{l|}{1.69$\rpm$0.28} &
  \multicolumn{1}{l|}{1.81$\rpm$0.46} &
  1.73$\rpm$0.46 &
  \multicolumn{1}{c|}{899.89$\rpm$1.90} &
  \textbf{1.07$\rpm$0.24} \\ \cline{2-8} 
\multicolumn{1}{|c|}{} &
  MultiVAE &
  \multicolumn{1}{l|}{2.23$\rpm$0.58} &
  \multicolumn{1}{l|}{2.91$\rpm$0.72} &
  \multicolumn{1}{l|}{3.55$\rpm$1.23} &
  2.98$\rpm$1.50 &
  \multicolumn{1}{c|}{771.26$\rpm$1.84} &
  \textbf{1.10$\rpm$0.39} \\ \cline{2-8} 
\multicolumn{1}{|c|}{} &
  NeuMF &
  \multicolumn{1}{l|}{0.82$\rpm$0.30} &
  \multicolumn{1}{l|}{0.85$\rpm$0.28} &
  \multicolumn{1}{l|}{1.51$\rpm$0.66} &
  1.69$\rpm$0.70 &
  \multicolumn{1}{c|}{758.45$\rpm$1.61} &
  \textbf{0.78$\rpm$0.27} \\ \cline{2-8} 
\multicolumn{1}{|c|}{} &
  itemKNN &
  \multicolumn{1}{l|}{1.84$\rpm$0.24} &
  \multicolumn{1}{l|}{2.13$\rpm$0.27} &
  \multicolumn{1}{l|}{1.97$\rpm$0.42} &
  2.17$\rpm$0.49 &
  \multicolumn{1}{c|}{725.72$\rpm$1.49} &
  \textbf{1.17$\rpm$0.28} \\ \cline{2-8} 
\multicolumn{1}{|c|}{} &
  ALS &
  \multicolumn{1}{l|}{9.41$\rpm$0.97} &
  \multicolumn{1}{l|}{12.83$\rpm$1.27} &
  \multicolumn{1}{l|}{10.63$\rpm$2.53} &
  10.57$\rpm$3.18 &
  \multicolumn{1}{c|}{705.76$\rpm$1.56} &
  \textbf{4.29$\rpm$1.05} \\ \hline
\end{tabular}%
}
\end{table*}

\end{document}